\documentclass[fleqn,usenatbib]{mnras}

\usepackage{newtxtext,newtxmath}
\usepackage{gensymb}
\usepackage{chngpage}
\usepackage{amsmath}

\usepackage[T1]{fontenc}
\usepackage{ae,aecompl}

\usepackage{graphicx}	
\usepackage{amsmath}	
\usepackage{amssymb}	

\usepackage{newtxtext,newtxmath}
\usepackage{gensymb}
\usepackage{amsmath}

\usepackage[T1]{fontenc}
\usepackage{ae,aecompl}

\usepackage{graphicx}	
\usepackage{amsmath}	
\usepackage{amssymb}	

\title[Gamma-ray emission from globular clusters]{Gamma-ray emission from high Galactic latitude globular clusters}

\author[S. J. Lloyd et al.]{
Sheridan J. Lloyd,$^{1}$\thanks{E-mail:sheridan.j.lloyd@durham.ac.uk (SJL)}
Paula M. Chadwick$^{1}$
and Anthony M. Brown$^{1}$
\\
$^{1}$ Centre for Advanced Instrumentation, Dept. of Physics, University of Durham, South Road, Durham, DH1 3LE, UK\\
\\
}

\date{Accepted 2018 August 6 . Received 2018 July 10; in original form 2018 March 15}

\pubyear{2018}

\begin{document}
\label{firstpage}
\pagerange{\pageref{firstpage}--\pageref{lastpage}}
\maketitle

\begin{abstract}
We analyse 8 years of \textsc{PASS} 8 \textit{Fermi}-LAT data, in the 60 MeV - 300 GeV energy range, from 30 \textcolor{black}{high Galactic latitude} globular clusters. Six of these globular clusters are detected with a TS > 25, \textcolor{black}{with NGC 6254 being detected as gamma-ray bright for the first time.} The most significant detection is of the well-known globular cluster 47 Tuc, and we produce a refined spectral fit for this object with a log parabola model. NGC 6093, NGC 6752 and NGC 6254 are fitted with hard, flat power law models, NGC 7078 is best fitted with a soft power law and NGC 6218 is best fitted with a hard, broken power law. This variety of spectral models suggests that there is a variety of $\gamma$-ray source types within globular clusters, \textcolor{black}{in addition to} the traditional millisecond pulsar interpretation. We identify a correspondence between diffuse X-ray emission in globular cluster cores and gamma-ray emission. This connection suggests that gamma-ray emission in globular clusters could \textcolor{black}{also} arise from unresolved X-ray sources or a relativistic electron population, perhaps generated by the millisecond pulsars. X-ray observations of further gamma-ray bright globular clusters would allow a functional relationship to be determined between diffuse X-ray and gamma-ray emission.       
\end{abstract}




\section{Introduction}

The 155 known (\cite{RN37}) globular clusters (GCs) are bound, spherical stellar systems. They are old  (of the order of 10\textsuperscript{10} years), dust-free satellites of the Milky Way galaxy, characterised by dense cores of 100 to 1000 stars per cubic parsec and consequently high stellar encounter rates. GCs are noted for hosting low mass X-ray binaries and populations of millisecond pulsars (MSPs) which arise from binary interactions. MSPs are strong gamma-ray sources emitting gamma-rays through curvature radiation and electron / positron pair production cascades in their magnetospheres. GCs may also contain central intermediate mass black holes (IMBH; \cite{RN124}) or reside within dark matter (DM) halos (\cite{1984ApJ...277..470P}), and in this context the annihilation of DM in IMBH could produce gamma-rays (\cite{RN180}). It is thus not surprising that GCs can also be significant gamma-ray sources. 

The \textit{Fermi}-LAT  has been surveying the entire gamma-ray sky in the energy range from 60 MeV to more than 300 GeV since its launch in 2008. \textit{Fermi}-LAT is a pair production instrument consisting of tracker, calorimeter and anti-coincidence detector modules. The tracker determines the direction of gamma-ray photons, the calorimeter measures the photon energy and the anti-coincidence detector vetoes false events caused by cosmic rays (\cite{RN195}). The LAT event level analysis framework has been refined since 2008 in successive data releases. \textsc{PASS} 6 was the initial data release, followed by \textsc{PASS} 7 in 2011, which improved the gamma-ray Galactic diffuse emission model, instrument response functions and direction reconstruction above 3 GeV. The latest data release is \textsc{PASS} 8, which is a complete reworking of the dataset which reduces gamma-ray background, increases instrument effective area, improves the point spread function and allows analysis down to 60 MeV.  

The first gamma-ray detected GC was 47 Tuc, with around 8 months of \textit{Fermi}-LAT (large area telescope) observations providing a detection with a significance of 17 $\sigma$ (\cite{RN196}). Later \textit{Fermi}-LAT observations show that 47 Tuc exhibits an exponential cutoff power law spectrum between 200 MeV and 10 GeV (\cite{RN196} and \cite{RN176}) while GC NGC 6093  (M80) has been identified as a possible gamma-ray source and a possible detection of NGC 6752  has been confirmed (\cite{RN50}). More recently NGC 6218 and NGC 7078 have been detected using \textit{Fermi}-LAT \textsc{PASS} 8 data (\cite{RN2}). 

However, the previous analyses of 47 Tuc, NGC 6093 and NGC 6752 were performed using just the 2 years of \textit{Fermi}-LAT \textsc{PASS} 6 data then available. Thus, the published SED for 47 Tuc is coarsely binned at 4 bins per decade of energy (\cite{RN176}), the possible detection of NGC 6093 has not been refined and no spectral energy distribution has been produced for NGC 6752 \cite{RN50}. 

To date 25 GCs are known gamma-ray emitters (\cite{RN56}) and a re-survey with the most up to date \textsc{PASS} 8 \textit{Fermi}-LAT data is likely to refine spectra further and possibly make fresh detections due to the 1-7 years of further photon statistics since the last publications and the improved effective area of the LAT instrument.  In addition, the latest \textsc{PASS} 8 data release and tools of the \textit{Fermi}-LAT, now allow spectral analysis in the 60 - 100 MeV range.  

This paper presents a new analysis of 30 GCs, and is structured as follows. In Section \ref{sec:GCSelection} we describe the selection criteria which result in the identification of the GCs for analysis. In Section \ref{sec:Analysis} we describe our analysis method for the detection of GCs, and variability, extension and spectral analyses for the GCs we detect. In Section \ref{sec:Results} we provide spectral models, spectra, light curves and flux determinations for the detected GCs along with their test statistic (TS) maps. For undetected GCs we present photon and energy flux upper limits. In Section \ref{sec:Discussion} we discuss whether the detected GC gamma-ray emission can be accounted for by MSPs on spectral grounds and determine a relationship between gamma-ray luminosity and diffuse X-ray luminosity in GCs. Finally in Section \ref{sec:Conclusion} we summarise our findings and make suggestions for future work.

\section{Globular Cluster Selection}
\label{sec:GCSelection}
Our selection of 30 GCs (Table~\ref{tab:gclist1_table}) is based on the following criteria to minimise background and facilitate interpretation of the results:
\begin{itemize}
\item Those GCs which are located off the Galactic plane (|b|> 15\textdegree{}) in order to mitigate Galactic gamma-ray background model uncertainty through the Galactic disc.
\item Those GCs which have a published mass (which is not simply an upper limit), surface brightness and absolute magnitude.  
\end{itemize}

This selection includes the previously detected GCs 47 Tuc, NGC 6093, NGC 6752 and NGC 7078.

\textcolor{black}{In order to exclude any selection bias caused by including only GCs with |b|> 15\textdegree{} we use the Kolmogorov-Smirnov (KS) test. We consider only those GCs with a helio-centric distance up to 10.4 kpc, the distance to NGC 7078, which is the furthest previously detected GC in our study. Our null hypothesis \textit{H\textsubscript{0}} is that the selected GCs in our study are consistent with the GCs in the Galactic plane for metallicity, mass and encounter rate. We individually test the  metallicity, mass and encounter rates of our study GCs against all other GCs with |b|< 15\textdegree{} using the KS test. We obtain KS test statistic (\textit{k}\footnote{\textit{k} = max(|F(x)\textsubscript{1}-F(x)\textsubscript{2}|) where F(x)\textsubscript{1,2} is proportion of x values less than current x for populations 1 and 2 being compared}) and probability (\textit{p}) values of 0.336/0.0512 (for metallicity), 0.233/0.376 (for mass) and 0.232/0.404 (for encounter rate). All determined probability values exceed the $\alpha$ significance level of 0.05 (which is the probability of falsely rejecting \textit{H\textsubscript{0}} when \textit{H\textsubscript{0}} is in fact true). Therefore we accept \textit{H\textsubscript{0}} that the study GCs metallicity, mass and encounter rate distribution are consistent with those of the GCs in the Galactic plane.}  

\begin{table*}
	\centering

	\begin{tabular}{lc c c c c c c c c lc lc} 
	
    \hline
Name&Helio&Metallicity&lii&bii&\textit{M\textsubscript{V}}&Core&Central&Mass&Mass&Gamma\\
&Distance/kpc&&&&&radius&surface& x 10\textsuperscript{5} M\textsubscript{$\odot$} &Ref&Source\\
&&&&&&&brightness&&&Ref\\
		\hline
NGC 6121&  2.2&-1.16&350.97&15.97& -7.19&1.16&17.95&1.01&[\citenum{RN135}]&\\
NGC 6752&  4.0&-1.54&336.49&-25.63& -7.73&0.17&14.88&1.4&[\citenum{SJL_1}]&[\citenum{RN50}]\\
NGC 6254&  4.4&-1.56&15.14&23.08& -7.48&0.77&17.7&2.26&[\citenum{RN135}]&\\
47 Tuc&  4.5&-0.72&305.89&-44.89& -9.42&0.36&14.38&7.0&[\citenum{SJL_1}]&[\citenum{RN176}]\\
NGC 6218&  4.8&-1.37&15.72&26.31& -7.31&0.79&18.1&1.44&[\citenum{RN42}]&[\citenum{RN2}]\\
NGC 6809&  5.4&-1.94&8.79&-23.27& -7.57&1.8&19.36&1.1&[\citenum{SJL_1}]&\\
NGC 6171&  6.4&-1.02&3.37&23.01& -7.12&0.56&18.94&0.96&[\citenum{RN135}]&\\
NGC 6205&  7.1&-1.53&59.01&40.91& -8.55&0.62&16.59&5.00&[\citenum{RN135}]&\\
NGC 6362&  7.6&-0.99&325.55&-17.57& -6.95&1.13&19.31&1.44&[\citenum{RN135}]&\\
NGC 7099&  8.1&-2.27&27.18&-46.84& -7.45&0.06&15.35&1.0&[\citenum{SJL_1}]&\\
E 3&  8.1&-0.83&292.27&-19.02& -4.12&1.87&23.1&0.14&[\citenum{RN177}]&\\
NGC 6341&  8.3&-2.31&68.34&34.86& -8.21&0.26&15.47&2.0&[\citenum{SJL_1}]&\\
NGC 362&  8.6&-1.26&301.53&-46.25& -8.43&0.18&14.8&3.21&[\citenum{RN135}]&\\
NGC 6723&  8.7&-1.1&0.07&-17.30& -7.83&0.83&18.13&1.96&[\citenum{RN135}]&\\
NGC 288&  8.9&-1.32&151.28&-89.38& -6.75&1.35&20.05&0.48&[\citenum{SJL_1}]&\\
NGC 6093& 10.0&-1.75&352.67&19.46& -8.23&0.15&15.11&3.37&[\citenum{RN135}]&[\citenum{RN50}]\\
NGC 5272& 10.2&-1.5&42.22&78.71& -8.88&0.37&16.64&5.00&[\citenum{RN135}]&\\
NGC 4590& 10.3&-2.23&299.63&36.05& -7.37&0.58&18.81&0.84&[\citenum{RN135}]&\\
NGC 7078& 10.4&-2.37&65.01&-27.31& -9.19&0.14&14.21&5.60&[\citenum{RN26}]&[\citenum{RN2}]\\
NGC 2298& 10.8&-1.92&245.63&-16.01& -6.31&0.31&18.9&0.56&[\citenum{RN42}]&\\
NGC 7089& 11.5&-1.65&53.37&-35.77& -9.03&0.32&15.78&7.64&[\citenum{RN135}]&\\
NGC 1851& 12.1&-1.18&244.51&-35.04& -8.33&0.09&14.25&2.99&[\citenum{RN135}]&\\
NGC 5897& 12.5&-1.9&342.95&30.29& -7.23&1.4&20.53&2.11&[\citenum{RN40}]&\\
NGC 1904& 12.9&-1.6&227.23&-29.35& -7.86&0.16&16.02&2.20&[\citenum{RN135}]&\\
NGC 5466& 16.0&-1.98&42.15&73.59& -6.98&1.43&21.61&1.04&[\citenum{RN42}]&\\
NGC 1261& 16.3&-1.27&270.54&-52.12& -7.80&0.35&17.73&2.23&[\citenum{RN42}]&\\
NGC 5053& 17.4&-2.27&335.70&78.95& -6.76&2.08&22.03&0.87&[\citenum{RN42}]&\\
NGC 5024& 17.9&-2.1&332.96&79.76& -8.71&0.35&17.38&3.83&[\citenum{RN135}]&\\
IC 4499& 18.8&-1.53&307.35&-20.47& -7.32&0.84&20.9&0.09&[\citenum{RN178}]&\\
NGC 4147& 19.3&-1.8&252.85&77.19& -6.17&0.09&17.38&0.53&[\citenum{RN135}]&\\
		\hline
	\end{tabular}
    	\caption{Selection of 30 GCs ordered by increasing distance from the Sun with name, helio distance (distance from Sun) in kpc and metallicity defined as [Fe/H]. lii and bii are Galactic longitude and latitude respectively in degrees, \textit{M\textsubscript{V}} is absolute visual magnitude, core radius is the radius of the GC core in arc mins and GC central surface brightness is in V Magnitudes/square arc sec from Harris 1996 and 2010. GC masses and previous identifications as a gamma-ray source are from the following references listed: [\citenum{RN176}]=\citeauthor{RN176}, [\citenum{RN135}]=\citeauthor{RN135}, [\citenum{SJL_1}]=\citeauthor{SJL_1}, [\citenum{RN40}]=\citeauthor{RN40}, [\citenum{RN178}]=\citeauthor{RN178}, [\citenum{RN26}]=\citeauthor{RN26}, [\citenum{RN177}]=\citeauthor{RN177}, [\citenum{RN50}]=\citeauthor{RN50}, [\citenum{RN42}]=\citeauthor{RN42}, [\citenum{RN2}]=\citeauthor{RN2} }
        \label{tab:gclist1_table}
\end{table*}

\section{Analysis}
\label{sec:Analysis}
\subsection{Photon Event Data Selection}
\label{sec:PhotonEventDataSelection} 
The data in this analysis were collected by \textit{Fermi}-LAT between 4th Aug 2008 to 28th December 2016 (Mission Elapsed Time (MET)  2395574147[s] to 504661408[s]). We consider all \textsc{pass} 8 events which are \textit{source} class  photons (evclass=128), both Front and Back converting events (evtype=3), spanning the energy range 60 MeV to 300 GeV. Throughout our analysis, the \textit{Fermipy} software package\footnote{\textit{Fermipy} change log version 0.12.0 (\cite{2017arXiv170709551W})} with version \textsc{v10r0p5} of the \textit{Fermi Science Tools} is used, in conjunction with the \textsc{p8r2\_source\_v6} instrument response functions. We apply the standard \textsc{pass}8 cuts to the data, including a zenith angle 90\textdegree{} cut to exclude photons from the Earth limb and good-time-interval cuts of DATA\_QUAL > 0 and LAT\_CONFIG = 1. The energy binning used is 8 bins per decade in energy and spatial binning is 0.1\textdegree{} per image pixel.

\subsection{Initial Detection of GC gamma-ray Emitters}
\label{sec:InititialDetectionGC} 
We firstly search for significant gamma-ray emitters from our list of GC targets. For each GC target, a 15\textdegree{} Radius of Interest (ROI) centred on the nominal GC co-ordinates is considered. The model we use in our likelihood analysis consists of a point source population seeded from the \textit{Fermi}-LAT's third point source catalog (3FGL), diffuse gamma-ray emission and extended gamma-ray sources. The diffuse emission detected by the \textit{Fermi}-LAT consists of two components: the Galactic diffuse flux, and the isotropic diffuse flux. The Galactic component is modelled with \textit{Fermi}-LAT's gll\_iem\_v06.fit spatial map with the normalisation free to vary. The isotropic diffuse emission is defined by \textit{Fermi's}  iso\_\textsc{P8R2}\_\textsc{SOURCE}\_\textsc{V6}.txt tabulated spectral data. The normalisation of the isotropic emission is left free to vary. 

We conduct an initial \textsc{binned} likelihood analysis, with the normalisation of all point sources within 15\textdegree{} of each GC target being left free, in addition to the spectral shape of all TS $>25$ sources. Point sources within the 10\textdegree{} to 25\textdegree{} from each GC target are frozen to their 3FGL values. From this initial likelihood fit, all point sources with a TS $<4$, or with a predicted number of photons, $Npred$, $<4$ are removed from the model. Thereafter a second likelihood fit is undertaken with this refined model. 

The best-fit model from this secondary likelihood fit is used with the \textit{Fermi Science Tool} \textsc{gttsmap}, to search for new point sources in the data that were not accounted for by the 3FGL. In particular, we then run \textit{Fermipy's `find\_sources'} method twice to detect all sources above 3$\sigma$ significance. \textit{Find\_sources} is a peak detection algorithm which analyses the test statistic (TS) map to find new sources over and above those defined in the 3FGL model by placing a test point source, defined as a power law (PL) with spectral index 2.0, at each pixel on the TS map and recomputing likelihood. We then run the \textit{optimize} method  which loops over all model  components in the ROI and fits their normalization and spectral shape parameters. It also computes the TS of all sources in the ROI. 

Sources with an offset less than 0.5\textdegree{} from the GC co-ordinates which are either unattributed point source detections or are a recognised GC with a 3FGL identifier are re-analysed using the same method, but with an expanded 25\textdegree{} ROI and 40\textdegree{} source region width in the energy range 60 MeV to 300 GeV. The results of this analysis are used for the variability  (Section \ref{sec:VariabilityAnalysis}) and GC extension (Section \ref{sec:ExtensionAnalysis}) analyses. 

We determine gamma-ray emission upper limits (UL) for undetected GCs by repeating the 100 MeV-300 GeV analysis as above and adding GC PL test point sources with \textit{index} 2.0, \textit{scale} 100 MeV and \textit{prefactor} = 1 x 10 \textsuperscript{-11} at the GC nominal co-ordinates after the \textit{setup} analysis step and before running \textit{find\_sources}.

\subsection{Variability}
\label{sec:VariabilityAnalysis}
The analysis output from \ref{sec:InititialDetectionGC} is used to construct a light curve for each gamma-ray bright GC by running the GTAnalysis \textit{lightcurve} method. \textit{lightcurve} fits the flux in a sequence of time bins by repeating the analysis steps of \ref{sec:InititialDetectionGC} for each time bin whilst freeing all spectral parameters of the GC and freezing all other source parameters. A bin size of 6 months is used because there are unlikely to be sufficient photon statistics to perform a good fit over a smaller time interval.

\subsection{Spatial Extension}
\label{sec:ExtensionAnalysis}

The analysis output from \ref{sec:InititialDetectionGC} is used to check for source extension for each detected GC by running the GTAnalysis \textit{extension} method. \textit{extension} replaces the GC point source spatial model with an azimuthally symmetric 2D Gaussian model. It then profiles likelihood with respect to spatial extension in a 1 dimensional scan to determine the likelihood of extension.

\subsection{Refining the GC Spectral Energy Distributions}
\label{sec:RefineGCSED} 
\textcolor{black}{We now optimise the spectral fits of detected GCs in light of the statistics obtained from the analysis in section \ref{sec:InititialDetectionGC}.} Globular clusters with a detection within 0.5\textdegree{} of the nominal GC co-ordinates are re-analysed between 100 MeV to 10 GeV with a 25\textdegree{} ROI and 40\textdegree{} source ROI width. This ROI and source ROI width exceeds that of the Fermi Science Support Center (FSSC) recommendation of 20\textdegree{}/30\textdegree{} respectively at 100 MeV and is used for consistency with the analysis in Section \ref{sec:Below100MeV}. 47 Tuc is analysed at 8 bins per decade of energy and NGC 6254 is analysed at 2 bins per decade, whilst the other GCs are analysed at 4 bins per decade, in order to ensure good photon statistics in each bin. All other criteria are as described above. Initially, the \textit{setup} and \textit{optimize} methods are run to create count and photon exposure maps and to compute the TS values of all 3FGL sources in the model.The \textit{fit} method are then run. \textit{fit} is a likelihood optimisation method which executes a fit of all parameters that are currently free in the the model and updates the TS and predicted count (npred) values of all sources. The normalisation of all sources within 10\textdegree{} of the GC are freed using  the \textit{free\_source} method to allow for the Point Spread Function (PSF) of front and back converting events down to 100 MeV . The source nearest to the GC centre had \textit{prefactor} and \textit{index} spectral parameters (Eqn.~\ref{PLeqn}) freed for power law sources, \textit{prefactor}, \textit{index1}, \textit{index2} (Eqn.~\ref{BKPLeqn}) freed for broken power power law sources and \textit{norm}, \textit{alpha} and \textit{beta} spectral parameters (Eqn.~\ref{LPeqn}) freed for a log parabola source. The shape and normalisation parameters of all sources with a TS > 25 are then individually fitted using the \textit{optimize} method. Finally, the \textit{fit} method is run twice more with an intervening \textit{find\_sources} step. The \textit{sed} method generated a spectral energy distribution, with energy dispersion disabled for GCs which are known 3FGL sources and a 5 $\sigma$ confidence limit on the determination of instrument upper limits.

\subsubsection{Searching for Emission below 100 MeV and above 10 GeV}
\label{sec:Below100MeV} 
Globular clusters with a detection within 0.5\textdegree{} of the nominal GC co-ordinates are re-analysed between 60 MeV to 300 GeV with a 25\textdegree{} ROI and 40\textdegree{} source ROI width. The ROI and source ROI width at 60 MeV are derived from extrapolation of the FSSC recommended ROI and ROI source width increase for that of a 1 GeV analysis going to 100 MeV  (15\textdegree{}/20\textdegree{} to 20\textdegree{}/30\textdegree{} respectively), as the worsening of the PSF from 100 MeV to 60 MeV is comparable to that between 1 GeV and 100 MeV. For front and back events combined, the 95 percent containment angle PSFs for energies of 1 GeV, 100 MeV and 60 Mev are 3\textdegree{}, 11\textdegree{} and 20\textdegree{} respectively. All other criteria and analysis steps are the same as Section \ref{sec:PhotonEventDataSelection} and Section \ref{sec:RefineGCSED} respectively. GCs which show emission below 100 MeV with all source normalisations freed within 10\textdegree{} of the GC are re-analysed with all source normalisations freed out to 20 \textdegree{} to allow for the 60 MeV PSF above. The results of that analysis are presented in \ref{sec:BELOW100}.

\subsubsection{Spectral Models}

The differential flux, \textit{dN/dE}, (photon flux per energy bin) of the detected GC is described through a power law (Eqn.~\ref{PLeqn}), broken power law (Eqn.~\ref{BKPLeqn}) or log parabola (Eqn.~\ref{LPeqn}) spectral model\footnote{As described in the FSSC source model \protect\cite{RN181}}. 

\begin{equation}{\label{PLeqn}}
\frac{dN}{dE}=N\textsubscript{0}\Big(\frac{E}{E\textsubscript{0}}\Big)^\gamma\
\end{equation}
where \textit{prefactor} = $N_{0}$, \textit{index}=$\gamma$ and \textit{scale}=$E_{0}$.

\begin{equation}{\label{BKPLeqn}}
  \frac{dN}{dE}=N\textsubscript{0}\times\begin{cases}
    (E/E\textsubscript{b})^{\gamma\textsubscript{1}}, & \text{if E<E\textsubscript{b}}.\\
    (E/E\textsubscript{b})^{\gamma\textsubscript{2}}, & \text{otherwise}.
  \end{cases}
\end{equation}
where \textit{prefactor} = $N_{0}$, \textit{index1}=$\gamma$\textsubscript{1}, \textit{index2}=$\gamma$\textsubscript{2} and \textit{breakvalue} = $E$\textsubscript{b}.

\begin{equation}{\label{LPeqn}}
\frac{dN}{dE}=N\textsubscript{0}\Big(\frac{E}{E\textsubscript{b}}\Big)^{-(\alpha+\beta \log ({E}/{E\textsubscript{b}}))}\
\end{equation}
where \textit{norm} = $N$\textsubscript{0}, \textit{alpha}=$\alpha$, \textit{beta}=$\beta$ and  $E$\textsubscript{b} is a \textit{scale} parameter.

In addition, all known sources take their 3FGL spectral shape.

The differential flux spectrum of millisecond pulsars is described by an exponential cut-off power law  (Eqn.~\ref{PLexp})

\begin{equation}{\label{PLexp}}
\frac{dN}{dE}=N\textsubscript{0}E\textsuperscript{-$\Gamma$}exp\Big(-\frac{E}{E\textsubscript{c}}\Big)
\end{equation}

where \textit{normalisation} = $N$\textsubscript{0}, \textit{spectral index}=$\Gamma$ and  $E$\textsubscript{c} is the \textit{cut-off} energy.

\section{Results}
\label{sec:Results}

\subsection{GC position, Emission and Model Parameters}
The source position, luminosity, energy flux, photon flux and spectral model parameters of the detected GCs (47 Tuc, NGC 6093, 6218, 6752 and 7078) are listed in Tables \ref{tab:gc_47_tuc_summary_table}, \ref{tab:gc_summary_table} and \ref{tab:gc_6218_summary_table}. The same parameters for the new detection of NGC 6254 are listed in Table \ref{tab:gc_summary_table} and also in Section \ref{sec:NGC6254}.  All values reported are for the analysis between 100 MeV and 10 GeV with 8 bins per decade in energy for 47 Tuc, 2 bins per decade for NGC 6254 and 4 bins per decade for all other detected GCs. We find that none of the GCs have significant emission above 10 GeV.

\subsection{47 Tuc (NGC 104)}

\begin{figure}
	\includegraphics[width=\columnwidth]{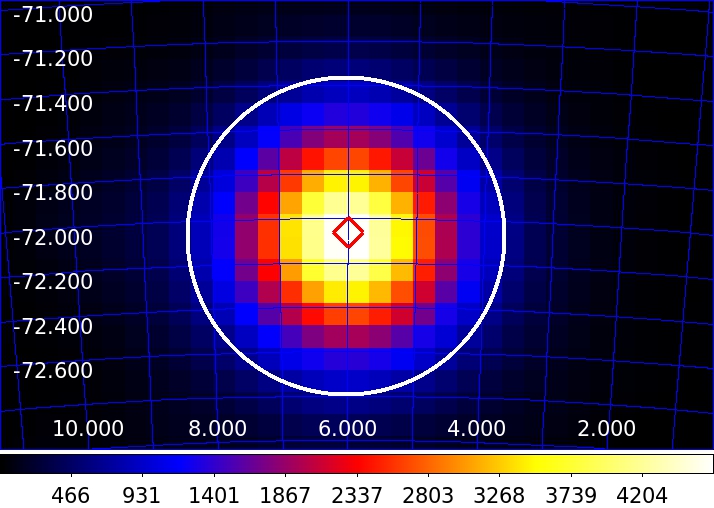}
    \caption{TS map of 47 Tuc with tidal radius 0.715\textdegree{} (white circle) and gamma-ray detection location (red diamond) marked. Graduated color bar (bottom) shows the TS value. RA and DEC are horizontal and vertical axes respectively on the white interior scale.}
    \label{fig:47_TUC_SIG}
\end{figure}

\begin{figure}
	\includegraphics[width=\columnwidth]{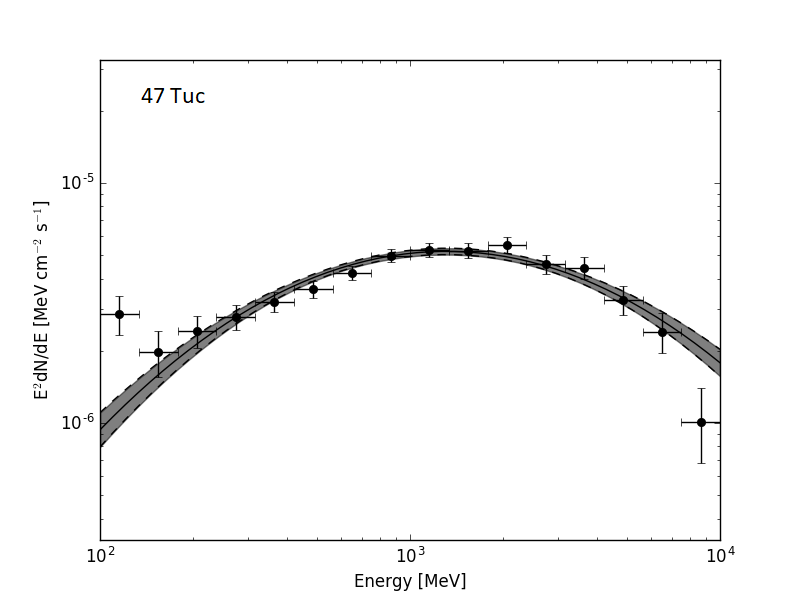}
    \caption{47 Tuc SED - The best fit (black line with uncertainty as grey band) is a log parabola as described in the FSSC source model \protect\cite{RN181} with spectral parameters \textit{norm} (7.0 $\pm$  0.2) x 10\textsuperscript{-12}, \textit{alpha} 1.67  $\pm$  0.04, \textit{beta} 0.38 $\pm$ 0.03 and Eb 856.5.}
    \label{fig:47_TUC_SED}
\end{figure}

47 Tuc is the second most massive GC in our sample after the undetected but more distant NGC 7089. Over the 8.3 year integration, 47 Tuc is detected with an overall significance of 72 $\sigma$ (TS 5229). 47 Tuc is associated with an existing 3FGL catalogue source 3FGL J0023.9-7203 which has a log parabola spectral model. Due to the longer exposure of our study, the gamma-ray spectrum of 47 Tuc (Fig.~\ref{fig:47_TUC_SED}) is refined at high and low energies compared to \cite{RN196} and \cite{RN176}. The best-fit log parabola model to this refined spectrum is also shown in Fig.~\ref{fig:47_TUC_SED} with the grey shaded region depicting the uncertainty associated with this best-fit. The spectral parameters of this best-fit model are listed in Table \ref{tab:gc_47_tuc_summary_table}. Interestingly, Fig.~\ref{fig:47_TUC_SED} shows tension between the observed spectrum and the best-fit model, particularly at low energies. The TS maps shows the detected source to be within the tidal radius of 47 Tuc (Fig. \ref{fig:47_TUC_SIG})

\begin{figure}
	\includegraphics[width=\columnwidth]{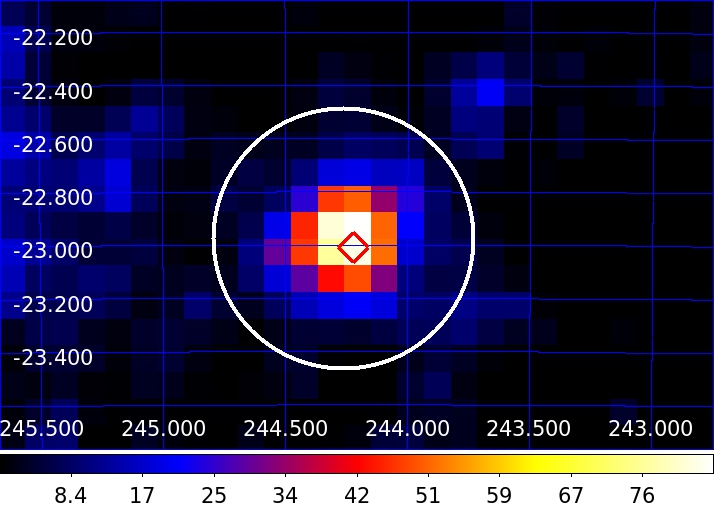}
    \caption{TS map of NGC 6093 with tidal radius 0.489\textdegree{} (white circle) and gamma-ray detection location (red diamond) marked. Graduated color bar (bottom) shows the TS value. RA and DEC are horizontal and vertical axes respectively on the white interior scale.}
    \label{fig:NGC_6093_SIG}
\end{figure}

\begin{figure}
	\includegraphics[width=\columnwidth]{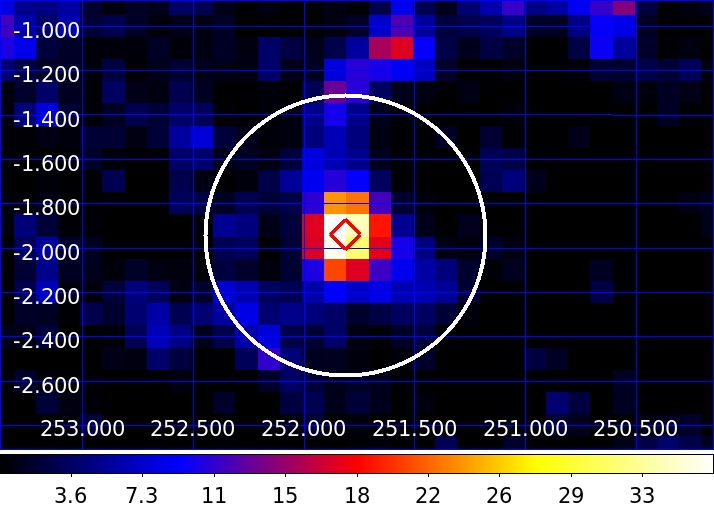}
    \caption{TS map of NGC 6218 with tidal radius 0.293\textdegree{} (white circle) and gamma-ray detection location (red diamond) marked. Graduated color bar (bottom) shows the TS value. RA and DEC are horizontal and vertical axes respectively on the white interior scale.}
    \label{fig:NGC_6218_SIG}
\end{figure}

\begin{figure}
	\includegraphics[width=\columnwidth]{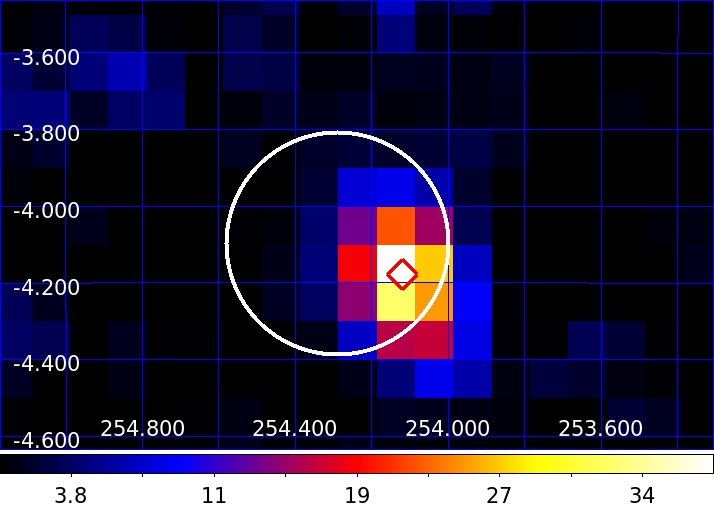}
    \caption{TS map of NGC 6254 with tidal radius 0.293\textdegree{} (white circle) and gamma-ray detection location (red diamond) marked. Graduated color bar (bottom) shows the TS value. RA and DEC are horizontal and vertical axes respectively on the white interior scale.}
    \label{fig:NGC_6254_SIG}
\end{figure}

\begin{figure}
	\includegraphics[width=\columnwidth]{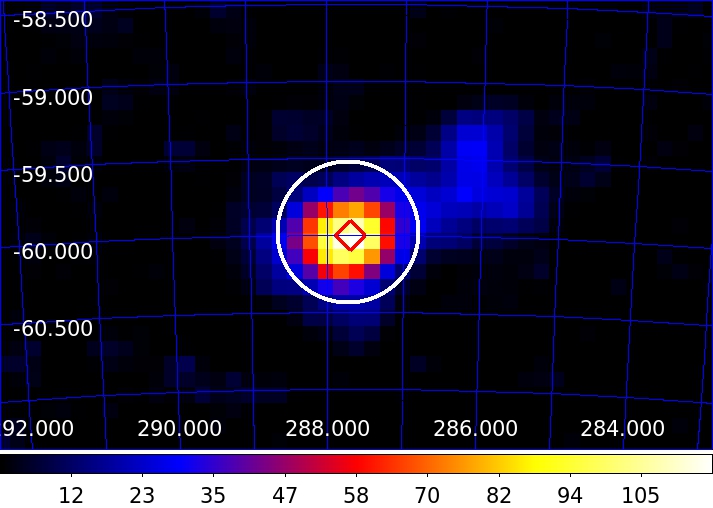}
    \caption{TS map of NGC 6752 with tidal radius 0.457\textdegree{} (white circle) and gamma-ray detection location (red diamond) marked. Graduated color bar (bottom) shows the TS value. RA and DEC are horizontal and vertical axes respectively on the white interior scale.}
    \label{fig:NGC_6752_SIG}
\end{figure}

\begin{figure}
	\includegraphics[width=\columnwidth]{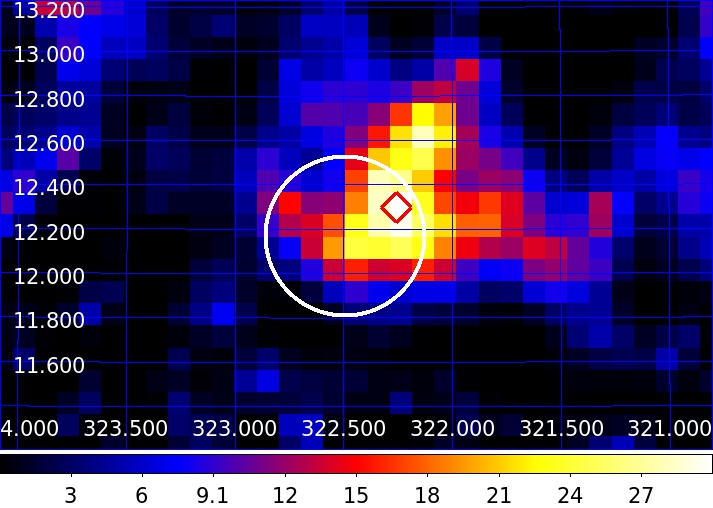}
    \caption{TS map of NGC 7078 with tidal radius 0.358\textdegree{} (white circle) and gamma-ray detection location (red diamond) marked. Graduated color bar (bottom) shows the TS value. RA and DEC are horizontal and vertical axes respectively on the white interior scale.}
    \label{fig:NGC_7078_SIG}
\end{figure}

\subsection{NGC 6093 (M80)}
\begin{figure}
	\includegraphics[width=\columnwidth]{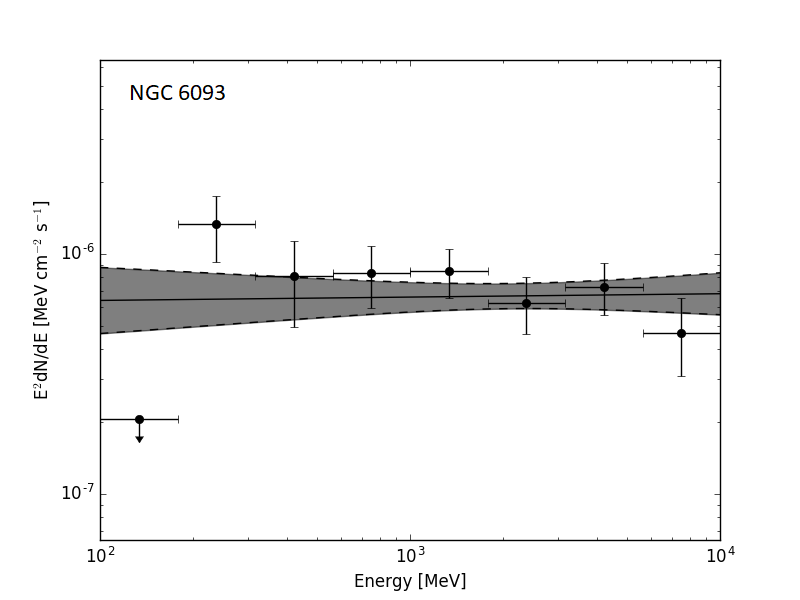}
    \caption{NGC 6093 SED fitted by a power law (black line with uncertainty as grey band) with functional form described in the FSSC source model (\protect\cite{RN181}). Spectral parameters are  \textit{prefactor}  (7.3 $\pm$  1.0) x  10\textsuperscript{-14}, \textit{index}  -2.13 $\pm$   -0.07 and \textit{scale} 2686}
    \label{fig:NGC_6093_SED}
\end{figure}

NGC 6093 is the seventh most massive GC in our sample (3.37 x 10 \textsuperscript{5} M\textsubscript{$\odot$}) and quite distant at 10.0 kpc. This GC is associated with the 3FGL source 3FGL J1616.8-2300. The possible detection of NGC 6093 at TS 27 (\cite{RN50}) is confirmed with an overall significance of 9.6 $\sigma$  (overall TS 94)  and a SED (Fig.~\ref{fig:NGC_6093_SED}) generated for the first time.  The emission is fitted by a power law (black line with uncertainty as grey band Fig.~\ref{fig:NGC_6093_SED}) with functional form described in the FSSC source model (\protect\cite{RN181}). The co-ordinates of the gamma-ray point source are RA = 244.24\textdegree{} and DEC =~-22.96\textdegree{}, which is within the tidal radius of the cluster on a TS map (Fig. \ref{fig:NGC_6093_SIG})

\subsection{NGC 6218 (M12)}

\begin{figure}
	\includegraphics[width=\columnwidth]{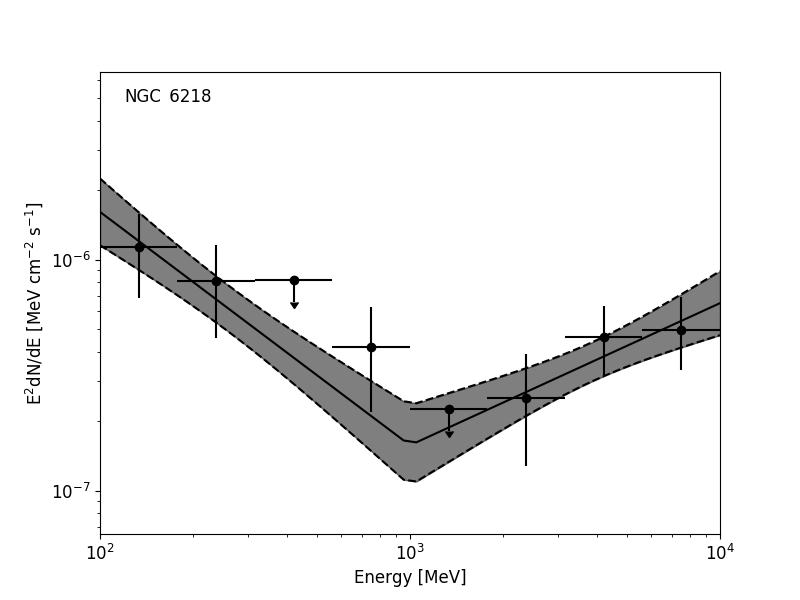}
    \caption{NGC 6218 SED fitted by a broken power law (black line with uncertainty as grey band) with functional form described in reference FSSC source model (\protect\cite{RN181}). Spectral parameters are \textit{prefactor} :(1.57 $\pm$  0.66) x 10 \textsuperscript{-13}, \textit{index1} -(3.00 $\pm$ 0.28), \textit{index2} -(1.39 $\pm$ 0.28) and \textit{scale}  1000}
    \label{fig:NGC_6218_SED}
\end{figure}

We refine the power law spectrum of \cite{RN2} for NGC 6218. Although the source can be fitted with a power law of overall significance of 6.4 $\sigma$ (TS 42), there is evidence that a broken power law with a break at 1 GeV is preferred over a simple power law (with a significance of 7.2 $\sigma$ (TS 52). (Fig.~\ref{fig:NGC_6218_SED}). The best fit position of the gamma-ray point source is RA = 251.82\textdegree{} and DEC = -1.90\textdegree{}, within the tidal radius of the cluster on a TS map (Fig. \ref{fig:NGC_6218_SIG}).

\subsection{NGC 6254 (M10)}
\label{sec:NGC6254}

\begin{figure}
	\includegraphics[width=\columnwidth]{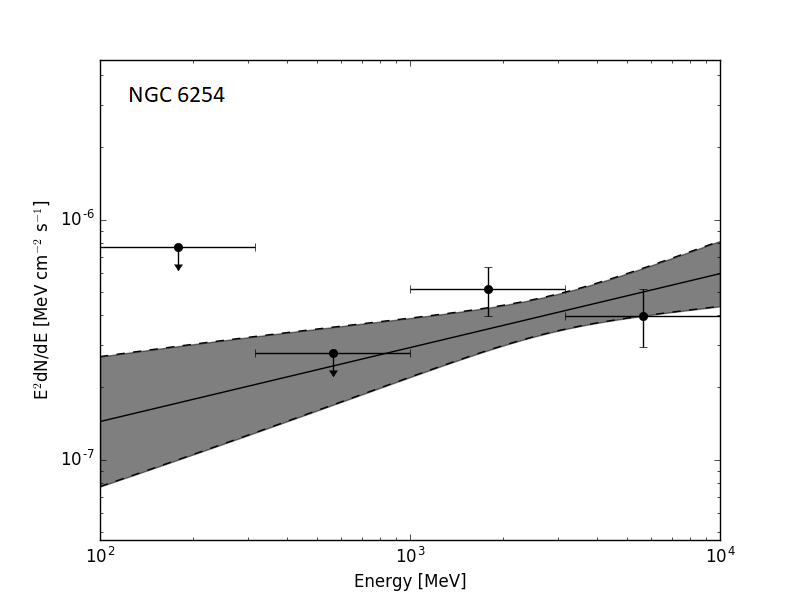}
    \caption{NGC 6254 SED fitted by a power law (black line with uncertainty as grey band) with spectral parameters: \textcolor{black}{\textit{prefactor} (2.9 $\pm$ 0.9) x 10\textsuperscript{-13}, \textit{index} -(1.69 $\pm$ 0.25), \textit{scale} 1000 }}
    \label{fig:NGC_6254_SED}
\end{figure}

NGC 6254 had not previously been detected with the \textit{Fermi}-LAT detector, and an upper energy flux limit of <2.14 x 10\textsuperscript{-12} erg cm\textsuperscript{-2} s\textsuperscript{-1} was given \cite{RN56}. The greater observation time used here provides a clear detection, with an overall TS of 40 (6.3 $\sigma$). The GC has an energy flux of (2.3 $\pm$ 0.5) x 10\textsuperscript{-12} erg cm\textsuperscript{-2} s\textsuperscript{-1} and a photon flux of (1.8 $\pm$ 1.1) x 10\textsuperscript{-9} cm\textsuperscript{-2} s\textsuperscript{-1}. The SED is fitted by a power law (black line with uncertainty as grey band Fig.~\ref{fig:NGC_6254_SED}) with a \textit{prefactor} of (2.9 $\pm$ 0.9) x 10\textsuperscript{-13}, \textit{index} -(1.69 $\pm$ 0.25) and \textit{scale} 1000. The co-ordinates of the gamma-ray point source are RA = 254.10\textdegree{} and DEC = -4.19\textdegree{}; although these are offset from the GC centre (RA = 254.2877, DEC = -4.1003) they are within the tidal radius 0.29\textdegree{} of the cluster on the TS map (Fig. \ref{fig:NGC_6254_SIG})


\subsection{NGC 6752}
\begin{figure}
	\includegraphics[width=\columnwidth]{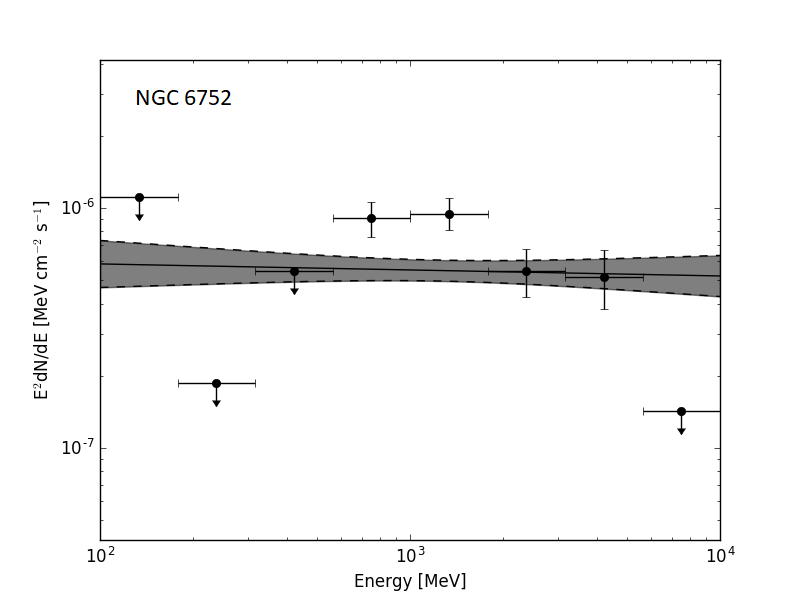}
    \caption{NGC 6752 SED fitted by a power law with spectral parameters \textit{prefactor} (4.0 $\pm$ 0.4) x 10\textsuperscript{-13}, \textit{index}  -(1.97 $\pm$  0.13), \textit{scale} 1272 }
    \label{fig:NGC_6752_SED}
\end{figure}

NGC 6752 was previously detected by \cite{RN50} in the energy range 200 MeV to 100 GeV. NGC 6752 is also associated with the 3FGL gamma-ray source 3FGL J1910.7-6000. The overall significance of the detection is 11.2 $\sigma$ (TS 126) with spatial offset 0.019\textdegree{}. We generate an SED for NGC 6752 for the first time (Fig.~\ref{fig:NGC_6752_SED}), which is fitted by a flat power law in the range 100 MeV  to 10 GeV. NGC 6752 shows some evidence for emission below 100 MeV as described in Section \ref{sec:BELOW100} The co-ordinates of the gamma-ray point source are RA = 287.72\textdegree{} and DEC = -60.02\textdegree{}, within the tidal radius of the GC (Fig. \ref{fig:NGC_6752_SIG}).

\subsection{NGC 7078 (M15)}
\begin{figure}
	\includegraphics[width=\columnwidth]{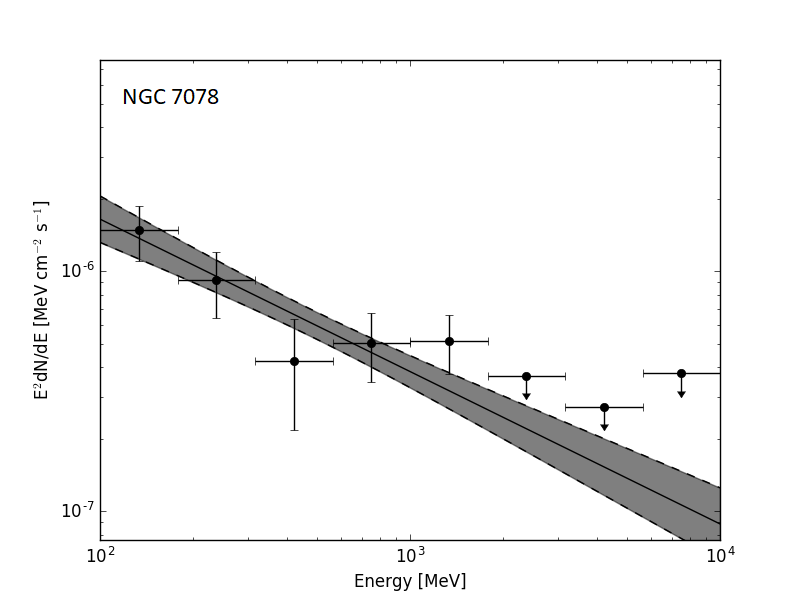}
    \caption{NGC 7078 - SED fit by a power law with spectral parameters: \textit{prefactor} (3.7 $\pm$ 0.6) x 10\textsuperscript{-13} , \textit{index} -(2.64 $\pm$ 0.13) , \textit{scale}  1000}
    \label{fig:NGC_7078_SED}
\end{figure}

NGC 7078 is the third most massive GC in our sample at 5.6 x 10 \textsuperscript{5} M\textsubscript{$\odot$} and quite distant at 10.4 kpc. We confirm the detection and SED of \cite{RN2} for NGC 7078, locating a point source, 0.266\textdegree{} from the GC co-ordinates with an overall significance 7.5 $\sigma$ (TS 56). The SED (Fig.~\ref{fig:NGC_7078_SED}) is fitted by a power law. The co-ordinates of the gamma-ray point source are RA = 322.29\textdegree{} and DEC = 12.32\textdegree{}, within the tidal radius of the cluster on a TS map (Fig. \ref{fig:NGC_7078_SIG}). The asymmetric TS map suggests that NGC 7078 could be comprised of multiple point sources. Attempting to resolve NGC 7078 into two point sources with an additional point source placed at a second significant point on the TS map (RA 322.20\textdegree{} and DEC 12.65\textdegree{}) does not yield a significant detection of the additional point source.

\subsection{Upper Limits}

The energy flux and photon flux upper limits for the 24 GCs in our sample which were not detected are presented in Table ~\ref{tab:ul_table}. 

\begin{table*}
	\centering
	
	\begin{tabular}{lc c lc lc lc } 
		\hline
                Globular & Energy flux UL  & Photon flux UL \\
                Cluster  & (10\textsuperscript{-13} erg cm\textsuperscript{-2} s\textsuperscript{-1}) &(10\textsuperscript{-10} cm\textsuperscript{-2} s \textsuperscript{-1})\\

        \hline
		\hline
E3 	        &			8.51				&	6.64		&\\
IC 4499 	&			19.43				&	15.15		&\\
NGC 288 	&			1.33				&	1.04		&\\
NGC 362 	&			19.91				&	15.53		&\\
NGC 1261 	&			1.47				&	1.14		&\\
NGC 1851 	&			19.93				&	15.54		&\\
NGC 1904 	&			23.75				&	18.52		&\\
NGC 2298 	&			5.77				&	4.50		&\\
NGC 4147 	&			2.54				&	1.98		&\\
NGC 4590 	&			4.48				&	3.50		&\\
NGC 5024 	&			10.55				&	8.23		&\\
NGC 5053 	&			13.80				&	10.76		&\\
NGC 5272 	&			16.20				&	12.63		&\\
NGC 5466 	&			8.03				&	6.26		&\\
NGC 5897 	&			12.40				&	9.66		&\\
NGC 6121 	&			47.57				&	37.09		&\\
NGC 6171 	&			16.73				&	13.05		&\\
NGC 6205 	&			14.13				&	11.02		&\\
NGC 6341 	&			23.08				&	18.00		&\\
NGC 6362 	&			3.04				&	2.37		&\\
NGC 6723 	&			18.80				&	14.66		&\\
NGC 6809 	&			17.35				&	13.53		&\\
NGC 7089 	&			3.73				&	2.91		&\\
NGC 7099 	&			6.51				&	5.08		&\\

		\hline
	\end{tabular}
    	\caption{Undetected GCs with upper limits (UL) for energy and photon flux at 95 percent confidence. The GCs are modelled as PL point sources with spectral \textit{index} 2.0 placed at the nominal GC co-ordinates  }
        \label{tab:ul_table}
\end{table*}

\subsection{Spatial Extension}

There is no evidence for spatial extension of gamma-ray emission in the detected GCs. The detected GCs are consistent with point-like gamma-ray emission sources.

\subsection{Variability}

\begin{figure}
	\includegraphics[width=\textwidth]{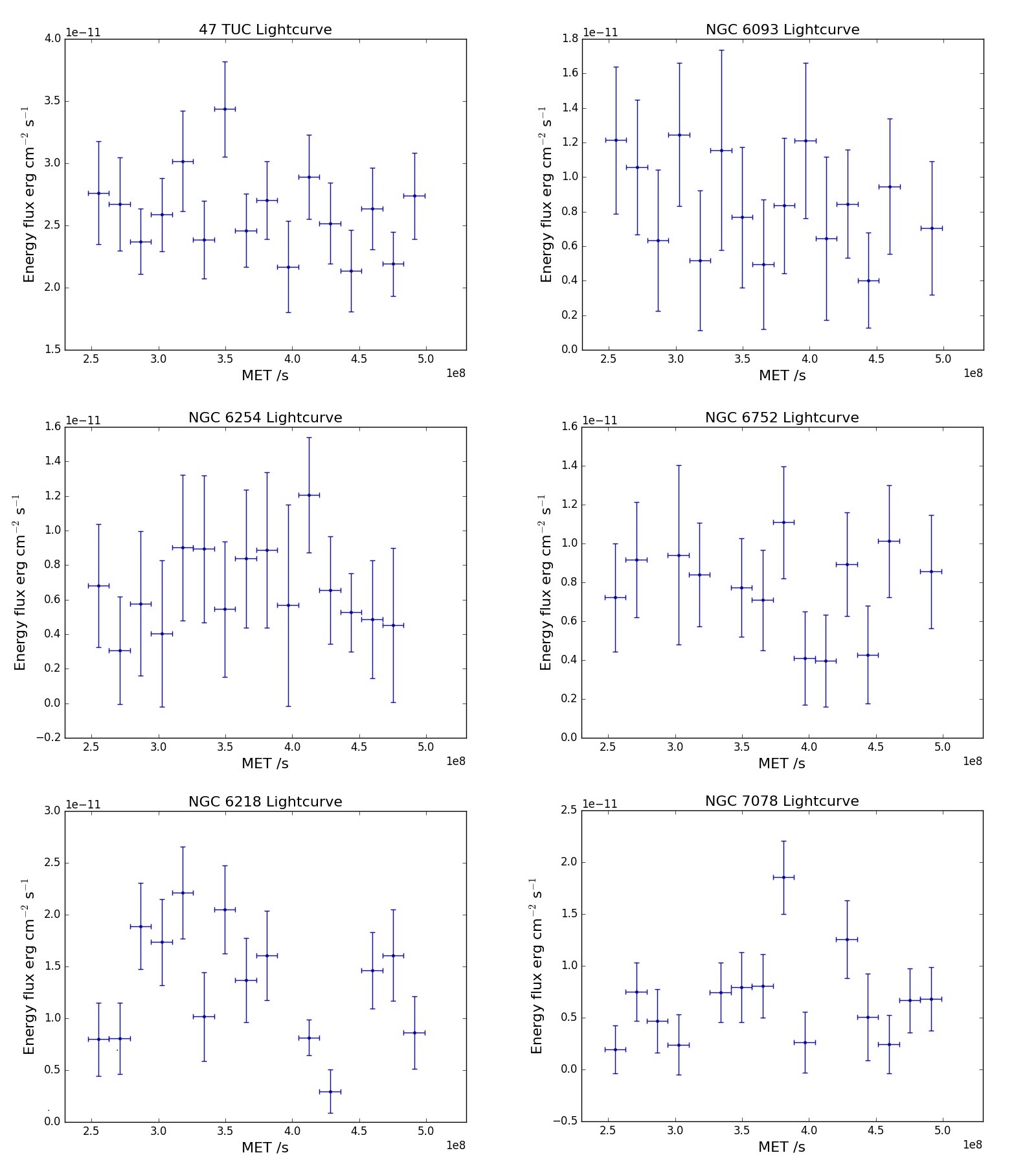}
    \caption{Lightcurves for the 6 detected GCs. Only NGC 6218 has some evidence of variability on 6 month timescales at the 3 $\sigma$ level }
    \label{fig:NON_VARIABLE}
\end{figure}

Light curves are generated for each detected GC in the range between 60 MeV to 300 GeV and are binned in time bins of 6 months (Fig.~\ref{fig:NON_VARIABLE}). The light curves for NGC 6093, NGC 6218, NGC 6752 and NGC 7078 show gaps in the binning where the optimisation and fitting process has not converged to an acceptable solution for the bin. A $\chi^2$ statistic is generated for a model comparing the observed fluxes in the 6 month bins against the average flux across all bins (Table~\ref{tab:chisquare_table}). The $\chi^2$ statistic exceeds the critical value at a probability of \textit{p}=0.999 for NGC 6218 ($\chi^2$ 40.52 vs critical value of 34.53) and \textit{p}=0.95 for NGC 7078. We therefore reject the null hypothesis of no significant variability in NGC 6218 at 3 $\sigma$ significance. This hypothesis is rejected for NGC 7078 at a lower significance of 2 $\sigma$, which is \textcolor{black}{insufficient} evidence for variability on a 6 month timescale. The $\chi^2$ statistic for 47 Tuc, NGC 6093, NGC 6254 and NGC 6752 is less than the critical value indicating no significant variability over a 6 month timescale at a probability of \textit{p}=0.95.

\begin{table*}
	\centering
	
	\begin{tabular}{lc c c c} 
		\hline
                Globular & $\chi^2$ & Degrees of & Upper Critical Value \\
                Cluster  & & freedom & \textit{p}=0.95\\        
		\hline
        \hline
        47 Tuc& 14.17 &     15&24.996   \\    
        NGC 6093& 7.49 &    14&23.685 \\
        NGC 6218& 40.52 &   13&22.363\\
        NGC 6254& 6.34 &    14&23.685\\
        NGC 6752& 9.24 &    12&21.026 \\
        NGC 7078& 24.97 &   13&22.362 \\
		\hline
	\end{tabular}
    	\caption{Detected GC $\chi^2$ values for a model comparing variable flux with average of variable flux across all bins. NGC 6218 and NGC 7078 exceed the critical value at \textit{p}=0.95 and thus show some evidence for variability on a 6 month time scale}
        \label{tab:chisquare_table}
\end{table*}

\subsection{Emission below 100 MeV}
\label{sec:BELOW100}
\begin{figure}
	\includegraphics[width=\columnwidth]{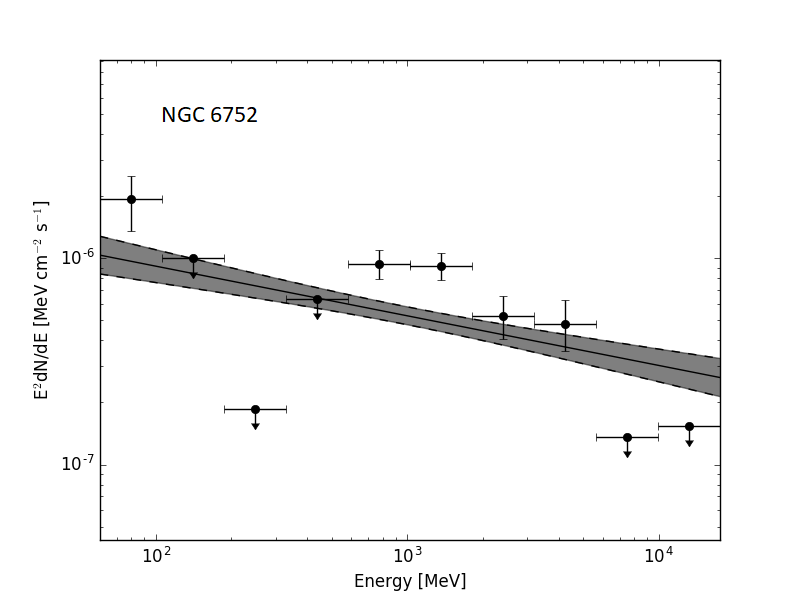}
    \caption{NGC 6752 - SED showing emission below 100 MeV. }
    \label{fig:NGC_6752_SED_LE}
\end{figure}

Only NGC 6752 shows any evidence for emission below 60 MeV. The sub 100 MeV flux is in the energy bin centered on 80 MeV and spanning 60 MeV to 106 MeV (Fig.~\ref{fig:NGC_6752_SED_LE}). This 80 MeV bin has an energy flux of 1.74 x 10\textsuperscript{-12} erg cm\textsuperscript{-2}  s\textsuperscript{-1} and a photon flux of 1.39 x 10\textsuperscript{-08} cm\textsuperscript{-2}  s\textsuperscript{-1}. The significance of the low energy flux bin is 3.4 $\sigma$ (TS 11.6).

\begin{table*}
	\centering
	
	\begin{tabular}{lc c lc lc  lc lc lc lc lc lc } 
		\hline
Globular&Source RA&Source DEC&Luminosity&Energy Flux&Photon Flux&\textit{norm}&$\alpha$&$\beta$&Eb\\
Cluster&Degree&Degree&10\textsuperscript{34} erg s\textsuperscript{-1}&10\textsuperscript{-11} erg cm\textsuperscript{-2}   \textsuperscript{-1}&10\textsuperscript{-8} cm\textsuperscript{-2} s\textsuperscript{-1}&10\textsuperscript{-12}&&&\\

        \hline
		\hline
47 Tuc&6.01 $\pm$ 0.01&-72.08 $\pm$ 0.01&6.29 $\pm$ 0.19&2.60 $\pm$ 0.08&2.45 $\pm$ 0.15&7.0 $\pm$ 0.2 &1.67  $\pm$  0.04&0.38$\pm$0.03&856.5\\

        \hline
	\end{tabular}
    	\caption{Detected GC 47 Tuc with positions (RA, DEC),luminosity, energy and photon flux and LP spectral model parameters of best fit on SED. Luminosity calculated assuming the distances in Table \ref{tab:gclist1_table}.}
        \label{tab:gc_47_tuc_summary_table}
\end{table*}

\begin{table*}
	\centering
	
	\begin{tabular}{lc c lc lc lc lc lc lc lc lc  } 
		\hline
Globular&Source RA&Source DEC&Luminosity&Energy Flux&Photon Flux&\textit{prefactor}&\textit{index}&\textit{scale}\\
Cluster&Degree&Degree&10\textsuperscript{34} erg s\textsuperscript{-1}&10\textsuperscript{-11} erg cm\textsuperscript{-2} s\textsuperscript{-1}&10\textsuperscript{-8} cm\textsuperscript{-2} s\textsuperscript{-1}&10\textsuperscript{-14}&&\\

        \hline
		\hline
NGC 6093&244.24 $\pm$ 0.02&-22.96 $\pm$ 0.02&5.84 $\pm$ 0.89&0.49 $\pm$ 0.07&0.64 $\pm$ 0.17&7.3 $\pm$ 1.0&-(2.13 $\pm$ 0.07)&2686\\
NGC 6254&254.10 $\pm$ 0.03&-4.19 $\pm$ 0.04&0.53 $\pm$  0.13&0.23 $\pm$ 0.05&0.18 $\pm$ 0.11&29 $\pm$ 9&-(1.69 $\pm$ 0.25)&1000\\
NGC 6752&287.72 $\pm$ 0.03&-60.02 $\pm$ 0.02&0.78 $\pm$ 0.08&0.41 $\pm$ 0.04&0.57 $\pm$ 0.10&40 $\pm$ 4&-(1.97 $\pm$ 0.13)&1272\\
NGC 7078&322.29 $\pm$ 0.07&12.32 $\pm$ 0.09&4.91 $\pm$ 0.73&0.50 $\pm$ 0.08&0.97 $\pm$ 0.19&37 $\pm$ 6&-(2.64 $\pm$ 0.13)&1000\\

		\hline
	\end{tabular}
    	\caption{Detected GCs with positions (RA, DEC),luminosity, energy and photon flux and PL spectral model parameters of best fit on SED. Luminosities calculated assuming the distances in Table \ref{tab:gclist1_table}. NGC 6254 is a new detection.  }
        \label{tab:gc_summary_table}
\end{table*}

\begin{table*}
	\centering
	
	\begin{tabular}{lc c lc lc lc lc lc lc lc lc  } 
		\hline
Globular&Source&Source&Luminosity&Energy Flux&Photon Flux&\textit{prefactor}&\textit{index1}&\textit{index2}&\textit{scale}\\
Cluster &RA Degree&DEC Degree&10\textsuperscript{34} erg s\textsuperscript{-1}&10\textsuperscript{-11} erg cm\textsuperscript{-2}   \textsuperscript{-1}&10\textsuperscript{-8} cm\textsuperscript{-2} s\textsuperscript{-1}&10\textsuperscript{-14}&&\\

        \hline
		\hline
         NGC 6218&251.82$\pm$0.03&-1.90$\pm$0.04 &0.99$\pm$0.17&0.36$\pm$0.07&0.82$\pm$ 0.24&15.7 $\pm$6.6&-(3.01$\pm$0.28)&-(1.39$\pm$0.28)&1000\\
		\hline
	\end{tabular}
    	\caption{Detected GC NGC 6218 with positions (RA, DEC),luminosity, energy and photon flux and Broken PL spectral model parameters of best fit on SED. Luminosity calculated assuming the distance in Table ~\ref{tab:gclist1_table}  }
        \label{tab:gc_6218_summary_table}
\end{table*}

\section{Discussion}
\label{sec:Discussion}

\subsection{Millisecond Pulsars and Globular Clusters: Spectral Characteristics}

\textcolor{black}{Before the launch of \textit{Fermi}, \cite{RN242} modeled the GC gamma-ray emission arising from the Comptonisation of stellar and CMB photons due to energetic leptons accelerated by millisecond pulsar wind shockwaves or from the magnetosphere. The spectra derived from this model predicted rising gamma-ray flux between 1-10 GeV and a hardening of the spectrum for 47 Tuc, NGC 7078 and NGC 6205, with the best candidates for gamma-ray detection predicted to be 47 Tuc and NGC 6205 (the latter being undetected in this study). \cite{RN248} predicted that curvature radiation (CR) (where relativistic electron / positron pairs are constrained to move along magnetic field lines) would produce gamma-ray emission peaking at 1-10 GeV from MSPs, while \cite{RN247} noted that the single particle CR spectrum of \cite{RN248} with its super exponential cut-off, would be reflected in the total phase averaged spectrum of an MSP with the cut-off arising from radiation reaction limited acceleration where acceleration rate of relativistic electrons equals the loss rate. This predicted spectral cut-off in MSPs was confirmed by \cite{RN82} who first observed gamma-ray pulsations from 8 MSPs (5 of which were in binary systems). They found that the gamma-ray spectrum of these MSPs was well described by a exponential cut-off power law (Eqn.~\ref{PLexp}) with a hard spectral index ($\Gamma$ < 2) and cut-off energy $E$\textsubscript{c} in the range 1-4 GeV. These spectral characteristics were found to hold more generally for a larger selection of 40 MSPs (10 isolated and 30 binary) in the second catalogue of \textit{Fermi}-LAT pulsars \cite{RN244}, hereafter the 2PC. Later, a stacked MSP spectrum was constructed from 39 of these 40 MSPs (10 isolated and 29 binary MSP systems) by \cite{RN197}, which showed a spectral cut-off at 5 GeV.}

\textcolor{black} {These average spectral characteristics of MSPs were used to identify the first gamma-ray emitting GCs. \cite{RN176} classified 5 gamma-ray sources as \textit{plausible} GC candidates on the basis of their spectral signature being MSP-like and matching that of the magnetospheric emission from an individual MSP with a spectral index < 2 and an exponential cut-off in the range 1.0-2.6 GeV. In contrast 3 sources were classed as \textit{possible} GCs because whilst they had a hard spectral index, they lacked evidence for an exponential cut-off in their spectra.}

\textcolor{black}{Observational evidence for the existence of individual MSPs (as opposed to an ensemble) within GCs may be found from radio observations. To date 150 pulsars have been detected and timed in 28 GCs, with the vast majority being MSPs\footnote{An list of currently known pulsars in GCs is maintained at http://www.naic.edu/{\textasciitilde}pfreire/GCpsr.html, accessed 2/7/2018}. Phase-resolved, pulsed gamma-ray emission from GCs is very rare, with pulses detected from a single gamma-ray bright MSP in only two GCs: NGC 6626 (\cite{RN240}) and NGC 6624 (\cite{RN33}). This provides an important link between GCs and gamma-ray emitting MSPs, albeit that these particular objects are unusually bright. MSP J1823-3021A in NGC 6624 has a gamma-ray luminosity of 8.4 x 10\textsuperscript{34} erg s\textsuperscript{-1} which can potentially account for the entire GC emission (\cite{RN33}) and MSP B1821-24 in NGC 6626 can account for 25 \% of the GC emission (\cite{RN240}). The gamma-ray spectra of these GC are fitted with a PL exponential model and have the spectral cut-offs characteristic of MSPs (1.5 GeV for NGC 6624 and 1-2.6 GeV for NGC 6626 respectively (\cite{RN50},\cite{RN176})). This confirms the view that the gamma-ray spectra of GCs, even when dominated by a small number of very bright MSPs, should exhibit spectral cut-offs and provides indirect support for the argument that an  ensemble of MSPs in GCs should also exhibit a spectral cutoff.}

\textcolor{black}{From the above we draw the following conclusions: the spectral characteristics of single MSPs are broadly predictive of the ensemble gamma-ray emission of MSPs in GCs in general, and the stacked spectrum of MSPs exhibits similar characteristics to the single MSP case. Furthermore, this stacked spectrum exhibits a spectral cut-off even when there is a mix of isolated and binary pulsar systems, as shown by \cite{RN244} and \cite{RN197}, where in fact binary systems were in the majority. Therefore, we take the characteristic spectral cut-off of single or stacked MSPs to be indicative of the expected spectra from MSPs in GCs. We will now consider the spectra of individual GCs in this context.}

\subsection{GC Spectral Shape and Potential Gamma-Ray sources}

\subsubsection{47 Tuc}

The 8.4 year spectrum of 47 Tuc is well described by a log-parabola model (Fig.~\ref{fig:47_TUC_SED}). However, there is tension between this best-fit model and the observed spectrum at the lowest and highest energies considered in this analysis. This tension may suggest that there are multiple emission sources within 47 Tuc. With 25 phase resolved MSPs, 47 Tuc has the second largest MSP population after that of Terzan 5 (\cite{RN215}). However, kinematic data has recently revealed possible evidence of an IMBH in 47 Tuc (\cite{RN124}). The presence of an IMBH within 47 Tuc raises the interestingly possibility of gamma-ray emission from DM annihilation (eg \cite{RN180}). Motivated by the possibility of an IMBH within 47 Tuc, a recent detailed spectral study found that the gamma-ray emission from 47 Tuc is best described by a two-source population model consisting of MSPs and annihilating DM, when compared to a MSP-only explanation (\cite{Brown}).

\subsubsection{NGC 6093, 6218 and 6254}

\begin{figure*}
	\includegraphics[width=\textwidth]{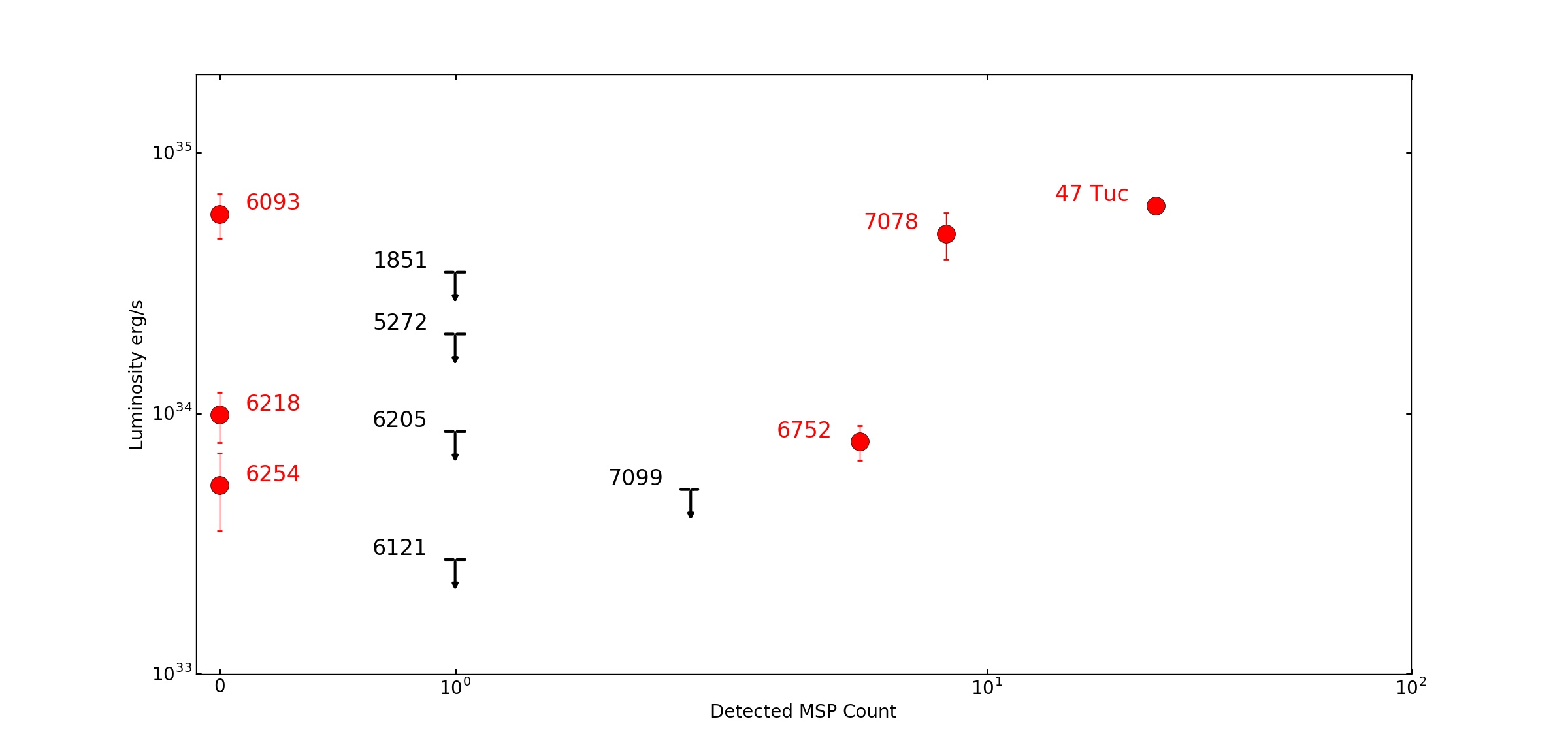}
    \caption{ Log Plot of gamma-ray luminosity vs number of known MSPs in each GC. We show GCs which are detected in this study (in red) and undetected GCs (in black) with known MSPs. For undetected GCs an upper luminosity limit is shown}
    \label{fig:MSP_COUNT}
\end{figure*}

The spectra of both NGC 6093 (Fig.~\ref{fig:NGC_6093_SED}) and NGC 6218 (Fig.~\ref{fig:NGC_6218_SED}) are hard, with NGC 6093 having a flat spectrum and NGC 6218 exhibiting increasing emission beyond $\sim$ 1 GeV. These GCs are not known to contain any MSPs (Fig ~\ref{fig:MSP_COUNT}), and it is perhaps not surprising that the spectra are unlike the typical stacked spectra of MSPs presented in \cite{RN197} where the flux falls 7 fold between 1-10 GeV.  Although the spectrum of the newly- (but also weakly-) detected GC NGC 6254 contains few points, this also appears to have a flat spectrum and, like NGC 6093 and NGC 6218, is not known to contain any MSPs (Fig ~\ref{fig:MSP_COUNT}). NGC 6218 shows some evidence for variability, which may point to a contribution from e.g. a background AGN. An X-ray study of NGC 6218 (\cite{0004-637X-705-1-175}) showed that there are several sources in the field of view, one of which, CX3, may be an AGN. The gamma-ray AGN catalogue is dominated by blazars - indeed, 98\% of the 3FGL AGN are this class of object (\cite{RN202}). There is no evidence that CX3 is a blazar, and given that blazars constitute only $\sim$~3\% of known AGN, the chance that it is a blazar is small. NGC 6218 also contains some CVs, but at this distance it is unlikely that they would be detected in gamma-rays.

\subsubsection{NGC 6752}

NGC 6752 has a hard, flat spectrum between 400 MeV and 4 GeV (Fig.~\ref{fig:NGC_6752_SED}), but a cut-off above 4 GeV. This object is known to contain 5 MSPs and the presence of this cut-off suggests that these are likely important contributors to the gamma-ray emission. 
In addition, there are 39 X-ray sources within the 1' 9" half-mass radius of NGC 6752, of which 16 are likely cataclysmic variables (CVs) and 3 are background AGN (\cite{RN125}). Three dwarf novae (CX1, CX4 and CX7) within this GC have been seen in outburst, over the last twenty years, using B band photometry and far UV/H$\alpha$ observations (\cite{RN192}, \cite{RN190}, \cite{RN125}). Some of these objects could be sources of gamma-ray emission, but the lack of gamma-ray variability on timescales of 6 months suggests this contribution is minor. 

\subsubsection{NGC 7078}

NGC 7078 (Fig.~\ref{fig:NGC_7078_SED}) has a soft power law spectrum, which is markedly different to the other GCs examined. The core of NGC 7078 was previously the target of a very long baseline interferometry radio study to constrain the mass of a putative central IMBH (\cite{RN122}). This found no evidence of central IMBH variability over a timescale of 2 months to 2 years, but did locate a strong radio source within 1.5 arc min of the GC centre. It was suggested that this radio source, S1, could be a background quasar. Similarly to NGC 6218, in the absence of a blazar classification (unlikely on population grounds), there is no evidence that S1 is the source of the gamma-ray emission. There are two further objects in the field of view of NGC 7078, both of which are low-mass X-ray binaries. None of this class of object is a confirmed gamma-ray source, so these also seem unlikely candidates for the gamma-ray emission. In the absence of any other plausible candidates, the working hypothesis is that the globular cluster is the source of the gamma-ray emission.

\subsubsection{Conclusion}

\textcolor{black}{The globular clusters 47 Tuc and NGC 6752 both show evidence of a spectral cut-off which could plausibly be explained by MSPs. The flat, hard spectra of NGC 6093, 6218 and 6254, which do not display a cut-off below 10 GeV are harder to explain by MSP emission, particularly given the the absence of radio-detected MSPs in these objects. NGC 6218's unusual spectrum is accompanied by evidence for variability at the $\sim~3\sigma$ level, which may point to a contribution from a variable source in the field of view, although a suitable candidate object appears lacking.}

\textcolor{black}{}

\subsection{Gamma-ray Luminosity and the Evidence for MSPs in GCs}

\textcolor{black}{The spectral evidence above hints at non MSP related gamma-ray sources in NGC 6093, 6218 and NGC 6254. Nonetheless, there are uncertainties in the precise spectrum that could arise from an ensemble of GC MSPs, so this evidence is not necessarily conclusive. However, if the gamma-ray emission for globular clusters derives exclusively or even primarily from MSPs, one would expect the gamma-ray luminosity to be correlated with the numbers of MSPs in the GCs in question. This presents its own problems, as the total number of MSPs in a given cluster is rarely (if ever) known.} 

\begin{figure*}
	\includegraphics[width=\textwidth]{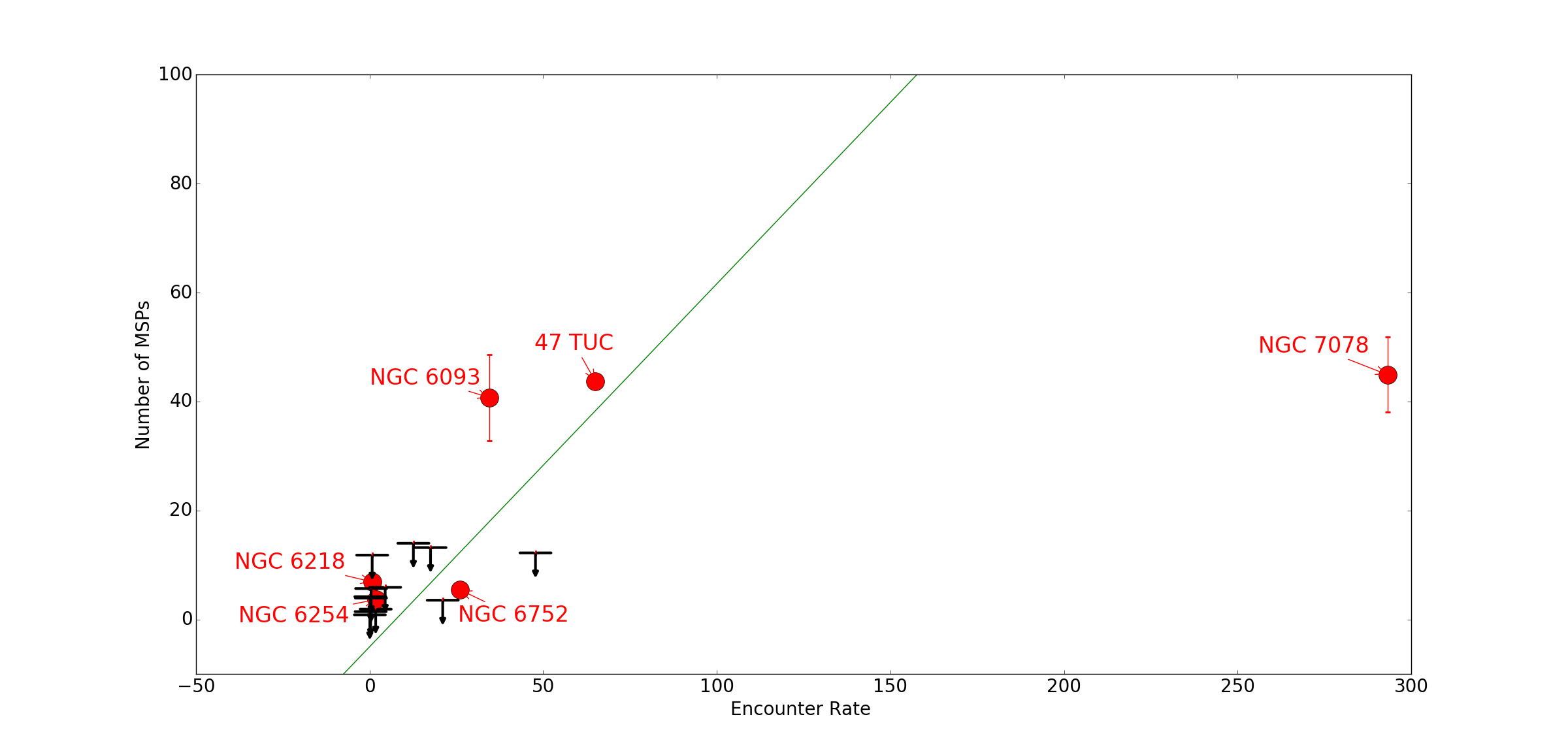}
    \caption{ \textcolor{black}{Plot of inferred MSP count vs Encounter Rate $\Gamma$\textsubscript{e} from \citeauthor{RN101} which we renormalise assuming 47 Tuc encounter rate is 65 to allow comparison with \citeauthor{RN176}. Detected GC are labeled (red captions) and we show upper limits (black symbols) only for those GC at a helio-centric distance of less than 10.4 kpc which is the furthest detection in our study (NGC 7078). The line of best fit (ignoring upper limit values) is shown in green and has the functional form  N\textsubscript{MSP}=(0.66$\pm$0.03)$\Gamma$\textsubscript{e} -(4.99$\pm$0.10)}  }
    \label{fig:REV_MSP_VS_ER}
\end{figure*}

\textcolor{black}{\cite{RN176} estimated the number of MSPs, N\textsubscript{MSP}, in 10 GCs from observed GC gamma-ray luminosity divided by expected emission for a canonical MSP (product of average E dot (1.1 x 10\textsuperscript{34} erg s\textsuperscript{-1}) and gamma-ray efficiency (0.08). They then derived a stellar encounter rate $\Gamma$\textsubscript{e} = $\rho$\textsubscript{0}\textsuperscript{1.5}r\textsubscript{c}\textsuperscript{2} for each GC, where $\rho$\textsubscript{0} is central cluster density in units of solar luminosity per pc\textsuperscript{3} and r\textsubscript{c} is cluster core radius in units of pc. The relationship between N\textsubscript{MSP} and $\Gamma$\textsubscript{e} was fitted by the linear relation    N\textsubscript{MSP}=(0.5$\pm$0.2)$\Gamma$\textsubscript{e} +(18$\pm$9). We determine N\textsubscript{MSP} as \cite{RN176} but use stellar encounter rates from \cite{RN101} which were derived using luminosity density profiles and velocity dispersion. We normalise the stellar encounter rates to \cite{RN176} values and plot N\textsubscript{MSP} vs stellar encounter rate Fig.~\ref{fig:REV_MSP_VS_ER} to yield the linear relationship N\textsubscript{MSP}=(0.66$\pm$0.03)$\Gamma$\textsubscript{e} -(4.99$\pm$0.10). Our result compares well with \cite{RN176} despite the only common detection being 47 Tuc with NGC 6752 and NGC 7078 (M15) presented only as upper limits in \cite{RN176}. This re-affirms connection between GC stellar encounter rate (which is presumed related to the number of MSPs) and gamma-ray luminosity. However, the detection of NGC 6218 and NGC 6254 (both with very low encounter rates and therefore presumably few, if any, MSPs) demonstrates that gamma-ray luminosity in GCs is not entirely related to the number of MSPs (Fig.~\ref{fig:REV_MSP_VS_ER}).}

\begin{figure*}
	\includegraphics[width=\textwidth]{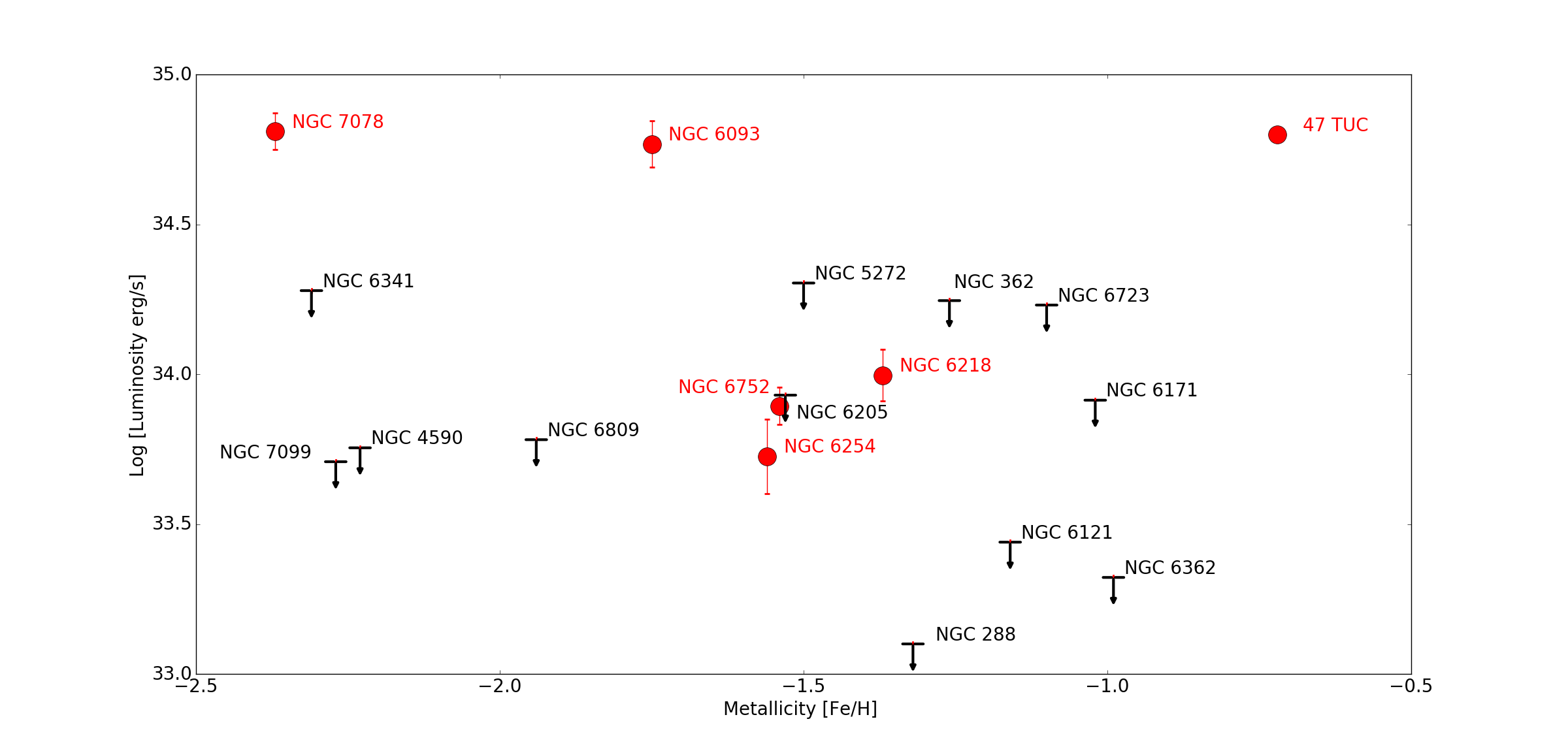}
    \caption{ \textcolor{black}{Plot of log gamma-ray luminosity vs Metallicity [Fe/H] for GCs with heliocentric distance of 10.4 kpc or less. Detected GCs are captioned in red whilst undetected GCs are captioned in black and shown as upper limits. There is no apparent correlation between luminosity and metallicity for the detected GCs in our study and NGC 6093 and NGC 7078 are the clear outliers. }}
    \label{fig:REV_METALLICITY}
\end{figure*}

\textcolor{black}{\cite{RN14} investigated the fundamental plane relations of gamma-ray globular clusters and determined a postive correlation between log of gamma-ray luminosity L\textsubscript{$\gamma$} and increasing metallicity [Fe/H] for 15 GCs including NGC 6752 and 47 Tuc. This linear relationship had the form  L\textsubscript{$\gamma$}=(0.59$\pm$0.15)[Fe/H] + (35.56$\pm$0.15 ). This correlation was ascribed to the increased likelihood of roche-lobe overflow and MSP recycling due to increased magnetic breaking in higher metallicity stellar systems as proposed by \cite{RN243}. We plot L\textsubscript{$\gamma$} vs metallicity for the detected GCs in our study and undetected GCs which are closer then the furthest detection (NGC 7078 at 10.4 kpc) but do not see an immediate correlation (Fig ~\ref{fig:REV_METALLICITY}) with NGC 7078 and NGC 6093 being the clear outliers. This reinforces the view that non MSP related sources of gamma-ray emission exist in NGC 6093 and NGC 7078.}

\begin{figure*}
	\includegraphics[width=\textwidth]{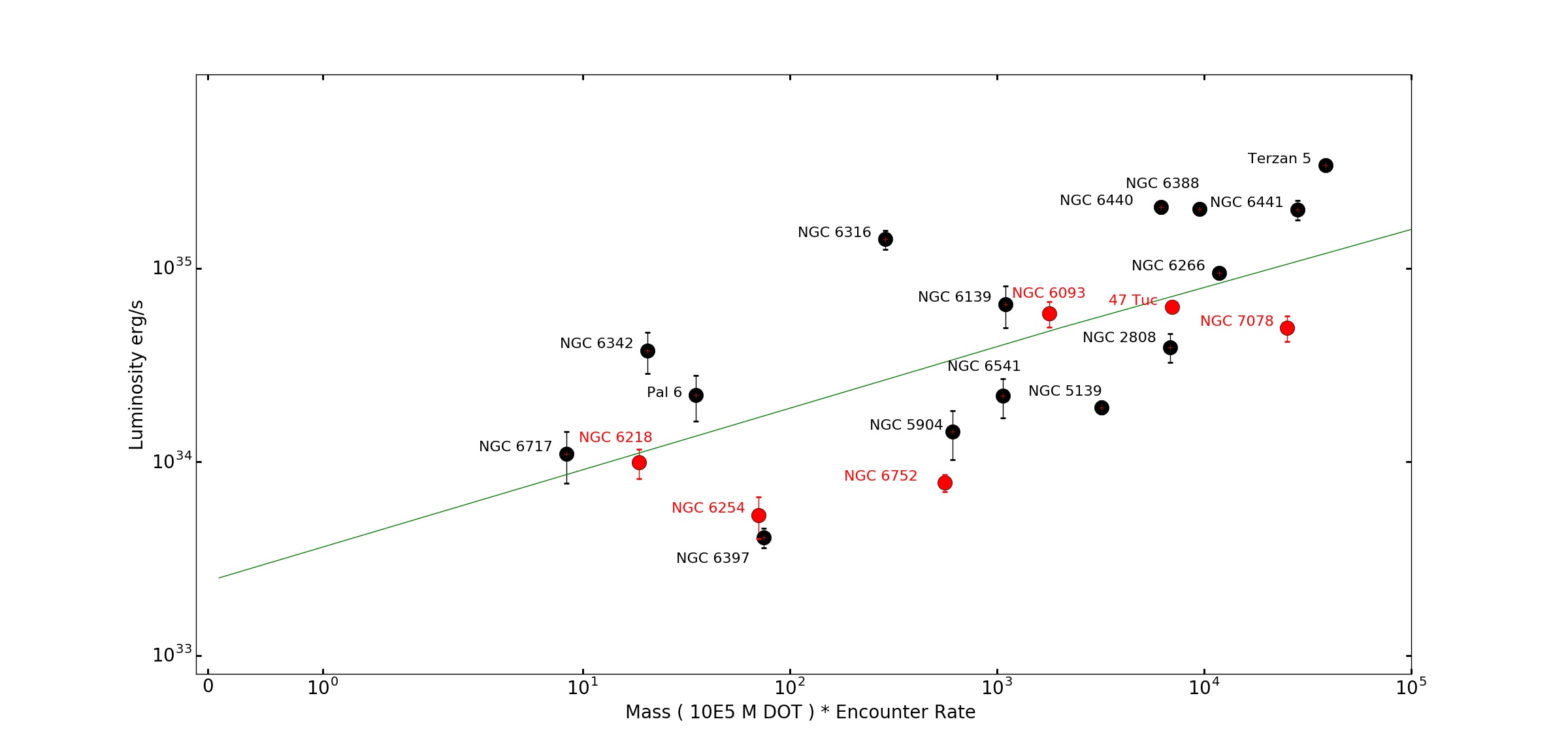}
    \caption{ Log plot of gamma-ray luminosity vs product of cluster mass and a normalised encounter rate from \citeauthor{RN101} for GCs detected in this study (in red) and selected GCs from \citeauthor{RN56} (in black) with known masses and encounter rates. In \citeauthor{RN101} the  encounter rate of 47 Tuc is set to an arbitrary value of 1000 and the encounter rate of other GCs are determined relative to that of 47 Tuc. The line of best fit (ignoring upper limit values) is shown in green and has functional form Log (L\textsubscript{$\gamma$})=0.30 Log (Mass*Encounter Rate) + 33.7 }
    \label{fig:ENCOUNTER_RATE_LUMINOSITY}
\end{figure*}

We \textcolor{black}{also} plot gamma-ray luminosity against \textit{detected} MSPs for the GCs in our study (Fig ~\ref{fig:MSP_COUNT}) and see that there is no correlation between detected MSP count and luminosity. For example, NGC 6093 and NGC 6218, with no detected MSPs, have comparable luminosity to 47 Tuc (25 MSPs) and NGC 6752 (5 MSPs) respectively. Therefore, to investigate further the connection between presumed binary-system creation and resulting MSPs and gamma-ray luminosity for the detected GCs in our sample and for a further 15 GCs in \cite{RN56} (with defined masses and encounter rates), we define a mass encounter rate product (MERP) derived from GC mass x GC normalised encounter rates (listed in \cite{RN101}) and plot against GC luminosity (Fig.~\ref{fig:ENCOUNTER_RATE_LUMINOSITY}). We take MERP as a proxy for the prevalence of binary system creation and MSP recycling.   

A line of best fit for Fig.~\ref{fig:ENCOUNTER_RATE_LUMINOSITY} determined by minimising  $\chi^2$ yields a tentative relationship between GC gamma-ray luminosity (L\textsubscript{$\gamma$}) and MERP (Eqn~\ref{MERPeqn}) which we plot on Fig.~\ref{fig:ENCOUNTER_RATE_LUMINOSITY}. Although the best-fit straight line of:

\begin{equation}{\label{MERPeqn}}
Log (L\textsubscript{$\gamma$})=0.30 Log (MERP) + \textcolor{black}{
33.7}
\end{equation}

is not compelling ($\chi^2$ = \textcolor{black}{3421/19} d.o.f.), it is preferred over a fit to constant average luminosity ($\chi^2$ = \textcolor{black}{39713/20} d.o.f.) 

There is therefore a weak correlation between gamma-ray luminosity and MERP for the 6 GCs we consider. This once again suggests that, while MSPs have a role to play in the gamma-ray emission from GCs, they are not necessarily the only source of the emission. We also note that mass alone appears unimportant; NGC 6093, which has a relatively low mass of 3.37 x 10\textsuperscript{5} M\textsubscript{$\odot$} and a distance of 10 kpc, is detected, whereas the larger clusters NGC 6205 and NGC 5272 (both of mass 5.00 x 10\textsuperscript{5} M\textsubscript{$\odot$} at distances of 7.1 kpc and 10.2 kpc respectively) are not.

\subsection{Diffuse Emission and Unresolved Point Sources}

 \textcolor{black}{The non MSP-like spectra of some our detected GCs coupled with the relatively poor correlation of gamma-ray luminosity  with metallicity or observed numbers of MSPs leads us to consider other sources of gamma-ray emission.} It is possible that the gamma-ray emission originates from as-yet unresolved point sources or has a more diffuse origin. We therefore explore the gamma-ray emission of GCs through its correlation with any diffuse X-ray emission excess, which is defined as the X-ray emission remaining after subtraction of all resolved X-ray point sources in a GC.
The spatial distribution of this diffuse emission is, however, complex, and may be split into two, largely separate, components: extended and core. The angular resolution of \textit{Fermi} is such that it is not possible to distinguish with which component the gamma-ray emission may be associated by positional coincidence.

\subsubsection{Extended X-ray Emission}

\textcolor{black}{Extended X-ray emission is defined as the observed diffuse X-ray excess in the band 0.5-4.5 keV, after the subtraction of known X-ray point sources, which extends beyond the GC core to the half-mass radius.}

\cite{RN146} used \textit{Chandra} to examine 12 GCs and detected extended X-ray emission in 6 objects, offset 1'-6' from the GC centre, associated with GC proper motion. For 5 GCs (47 Tuc, NGC 6752, 5904, 6093 and 6266) this emission was ascribed to the GC motion through the Galactic halo and shock heating of internal gas interacting with halo plasma, whereas in the case of GC Omega Centauri (NGC 5139) a background cluster of galaxies was the likely X-ray source. Subsequently \cite{RN204}, using \textit{Suzaku}-XIS, ascribed the putative 47 Tuc extended diffuse X-ray emission to a background cluster of galaxies with redshift \textit{z}=0.34.

In contrast, \cite{RN133} conducted a search for extended X-ray emission in 6 \textit{Fermi}-LAT detected GCs (NGC 6266, 6388, 6541, 6626, 6093 and 6139) in concentric zones between half to 4 times the GC half mass radius. They concluded that there was no evidence for diffuse emission above the level of the Galactic diffuse background in these GCs. The lack of extended X-ray emission in  \textcolor{black}{NGC 6093 and NGC 6266} is at odds with the above detections of \cite{RN146}. This is likely due to the different methods used by the authors to determine and account for the Galactic diffuse X-ray background. \cite{RN133} model the Galactic diffuse X-ray background by scaling the flux measurements of \cite{RN194} using expected X-ray emission and absorption arising from the GC interstellar medium.  In contrast, \cite{RN146} subtract an exposure corrected observational background, a few arc-minutes from the aim point.

\subsubsection{Core X-ray Emission}
 
 \textcolor{black}{Core X-ray emission is defined as the observed diffuse X-ray excess in the band 0.3-8 keV, after the subtraction of known X-ray point sources within the GC core radius.} \cite{RN147} examined the unresolved core X-ray emission of 10 MSP-hosting GCs through subtraction of known X-ray point sources detected with the \textit{Chandra} X-ray observatory. Diffuse X-ray core emission was detected in 4 GCs (NGC 6626, 6440, 6266 and 6752), which was then fitted with power law (PL) and thermal Bremsstrahlung (BREMSS) models. They linked this unresolved X-ray emission to the cumulative contribution of CVs and faint MSPs.  \cite{RN203}, again using \textit{Chandra}, identified diffuse X-ray emission from 47 Tuc consisting of 2 components, one a non-thermal component correlated with the GC stellar density profile, and the other a uniform thermal component offset from the GC. They interpreted the non-thermal X-ray emission from the core as resulting from shock fronts of stellar winds and inverse-Compton scattering of relic photons by pulsar winds. In their study, which largely concentrated on extended emission, \cite{RN146} noted that NGC 5904 (M5) showed evidence for two X-ray components: a pair of soft, wing-like regions and harder emission associated with the core. The latter was also interpreted as arising from an ensemble of faint point sources. However, in the case of Terzan 5, \cite{RN193} note that the contribution of unresolved sources to the centrally-peaked X-ray emission which they identified in \textit{Chandra} data is likely to be negligible.

\subsubsection{X-ray and $\gamma$-ray Emission}

We summarise this information, for GCs from this study and others for which information is available regarding their diffuse X-ray emission, in Table \ref{tab:lit_gc_yes}. In so far as there is  \textcolor{black}{an association}  between diffuse X-ray emission and gamma-ray detection, it appears that the objects with core X-ray emission are more likely to be gamma-ray emitters than not, and that the presence of extended X-ray emission is not as important. It therefore seems likely that the gamma-ray emission arises from the cores of the globular clusters rather than any extended region.

The question then arises as to the source(s) of the gamma-ray emission. We have noted there appears to be no strong connection with the number of MSPs known to exist in the objects, although there does appear to be some correlation with the mass $\times$ normalised encounter rate of the GCs. Assuming there is a connection to the core X-ray emission, it is not clear whether this arises from several unresolved point sources or genuinely diffuse emission. 

If the gamma-ray emission comes from unresolved point sources other than MSPs, the obvious candidates are cataclysmic variables (CVs) and X-ray binaries (XRBs), both of which are expected to exist within GCs due to the high stellar encounter rates. While gamma-ray emission has been detected from cataclysmic variables (\cite{RN246}), the emission is transient in nature, with gamma-ray emission only observed on the timescales of days after the novae event. Furthermore, only CVs in our local Galactic neighbourhood have been observed to be gamma-ray bright, which suggests that their gamma-ray emission is detected primarily because of their proximity (\cite{doi:10.1093/mnras/stw2776}). X-ray binaries are a less likely prospect, as XRBs in GCs will be low-mass systems, whereas nearly all the known gamma-ray emitting XRBs are wind-driven, high-mass systems, the one possible exception being XSS J12270-4859 (\cite{refId0}).

One potential source of the diffuse X-ray emission in these objects is relativistic electrons provided by the MSP population which can produce X-ray emission via synchrotron radiation. As both \cite{RN146} and \cite{RN193} have pointed out, assuming a typical Galactic magnetic field of a few $\mu$G, the electrons would require an energy of $\sim 10^{14}$~eV to produce emission at keV energies. Associated TeV gamma rays produced via inverse Compton radiation would be diagnostic of the existence of relativistic electrons and a low magnetic field. In this context, we note that Terzan 5 has been detected at TeV energies with the H.E.S.S. telescopes (\cite{2011A&A...531L..18H}), but that this seems to be a unique object, with TeV upper limits being obtained for several other GCs, including 47 Tuc and NGC 7078 (\cite{2013A&A...551A..26H}). GCs could also be an interesting prospect for the forthcoming Cherenkov Telescope Array (\cite{ACHARYA20133}).

Finally, the presence of IMBHs in some GCs (\cite{RN124}) could indicate a dark matter annihilation component to the emission in the more massive objects.

\begin{table*}
	\centering

	\begin{tabular}{l c c  c c c c c} 
		\hline
                    GC Name                             &L\textsubscript{$\gamma$}&Core Diffuse             &Core PL             &Core BREMSS         &Extended Diffuse  &Shock Front \\
                    and L\textsubscript{$\gamma$} Ref           &10\textsuperscript{34} erg s\textsuperscript{-1}   &X-ray            &L\textsubscript{x} 0.3-8.0 keV      &L\textsubscript{x} 0.3-8.0 keV        & X-ray emission&L\textsubscript{x} 0.5-4.5 keV \\
                    
                               &&    emission           &10\textsuperscript{32} erg s\textsuperscript{-1}  &10\textsuperscript{32} erg s\textsuperscript{-1}        &  &10\textsuperscript{32} erg s\textsuperscript{-1}\\

		\hline
                    
                    \textbf{GC detected}&   && & &&\\
                    \textbf{in {$\gamma$}-ray this work:}&   && & &&\\
                    47 TUC&6.29 $\pm$ 0.19   &Yes [\citenum{RN203}]                    &   $2.36\substack{+0.50 \\ -0.48}$ &     -     &Yes [\citenum{RN203}]&$1.26\substack{+0.07 \\ -0.08}$ \\
                    NGC 6093&5.84 $\pm$ 0.89 &-                    &    -  &     -     &Yes [\citenum{RN146}]&0.36$\pm$0.24\\
                    NGC 6752&0.78 $\pm$ 0.08 &Yes [\citenum{RN147}]&$0.15\substack{+0.08 \\ -0.04}$ &$0.12\substack{+0.11 \\ -0.06}$          &Yes [\citenum{RN146}]&6.0$\pm$0.3\\
                    \hline
                    \textbf{GC undetected}&        && & &&\\
                    \textbf{in {$\gamma$}-ray this work:}&   && & &&\\
                    NGC 5024&<4.04 &No   [\citenum{RN147}]&-&-&-                   &- \\
                    NGC 5272&<2.02 &No   [\citenum{RN147}]&-&-&-                   &- \\
                    NGC 6121&<0.27 &No   [\citenum{RN147}]&-&-&No [\citenum{RN146}]& <0.04\\
                    NGC 6205&<0.85 &No   [\citenum{RN147}]&-&-&-                   &- \\
                    NGC 7099&<0.51 &-&-&-                 &No [\citenum{RN146}]& <0.21\\
		\hline

		\hline
                    \textbf{GCs in literature}&        && & &&\\
                    NGC 5139 [\citenum{RN176}]   &2.8$\pm$0.7 & -                       &- &-                                                     &Yes - Galaxy cluster source [\citenum{RN146}]&  -\\
                    NGC 5904 [\citenum{RN2}]   &2.18$\pm$0.4   & No  [\citenum{RN147}] / Yes [\citenum{RN146}]  &M5C 0.8$\pm$0.2&-&Yes [\citenum{RN146}]&M5W 0.1$\pm$0.03 \\
                    NGC 6266 [\citenum{RN176}]   &$10.9\substack{+3.5 \\ -2.3}$ &Yes [\citenum{RN147}]    & $1.5\substack{+0.62 \\ -0.36}$  & $1.41\substack{+0.26 \\ -0.27}$                                                     &Yes [\citenum{RN146}] & 0.6$\pm$0.2&\\
                    NGC 6397 [\citenum{RN2}]   &$0.2\substack{+0.15 \\ -0.12}$ &-    &-  & -&No [\citenum{RN146}] & <0.13&\\

                    NGC 6440 [\citenum{RN176}]   &$19.0\substack{+13.1 \\ -5.0}$ &Yes [\citenum{RN147}]    &$2.00\substack{+1.71 \\ -0.76}$  &$1.57\substack{+1.51 \\ -0.74}$ &No  [\citenum{RN146}]&<0.19\\
                    NGC 6626 [\citenum{RN176}]   &$6.2\substack{+2.6 \\ -1.8}$ & Yes [\citenum{RN147}]   &$2.54\substack{+1.27 \\ -0.62}$   & $1.67\substack{+0.99 \\ -0.62}$ &No  [\citenum{RN146}]&<0.51\\
                    Terzan 5     [\citenum{RN176}]   &$25.7\substack{+9.4 \\ -8.8}$ & Yes [\citenum{RN193}]   &20$\pm$3 (1-7 keV)  &-      &-&-\\
                    NGC 6366 [\citenum{RN56}]  &<0.47 & -                     &-&-&No [\citenum{RN146}]&<0.62\\
                    NGC 6838 [\citenum{RN56}]  &<1.14 & No  [\citenum{RN147}] &-&-&-&-\\

		\hline
	\end{tabular}
    	\caption{Correspondence of detected diffuse X-ray and  gamma-ray emission for GCs.  47 Tuc, NGC 6093 and NGC 6752 have diffuse X-ray emission and are detected in gamma-ray whereas GCs 5024 to NGC 7099 lack diffuse X-ray emission and are not detected in gamma-ray. Core X-ray emission is determined either as power law (PL) or thermal Bremsstrahlung component (BREMSS) and GCs for which the diffuse X-ray emission is undetermined are indicated with a "-"  NGC 5904 has two distinct X-ray emitting regions M5C (core) and M5W (wing) both observed between 0.5-4.5 keV. NGC 6397 is a marginal case with low gamma-ray luminosity and X-ray upper limit. Gamma-ray and X-ray luminosities from references listed: [\citenum{RN176}]=\citeauthor{RN176}, [\citenum{RN193}]=\citeauthor{RN193}, [\citenum{RN56}]=\citeauthor{RN56}, [\citenum{RN147}]=\citeauthor{RN147},  [\citenum{RN146}]=\citeauthor{RN146},[\citenum{RN203}]=\citeauthor{RN203} and [\citenum{RN2}]=\citeauthor{RN2} X-ray energy bands as indicated in table except for 47 Tuc observed between 0.5-7 keV 
}
        \label{tab:lit_gc_yes}
\end{table*}

\section{Conclusions}
\label{sec:Conclusion}

We analyse 8 years of \textsc{PASS} 8 \textit{Fermi}-LAT data from 30 globular clusters. We refine the gamma-ray spectra of 5 previously detected GCs and detect NGC 6254 for the first time. NGC 6752 lacks detectable gamma-ray emission above 4 GeV, suggesting that this is the one object with emission dominated by MSPs. However, the spectral shapes of NGC 6093, NGC 6254, 47 Tuc and NGC 6218 suggest that other sources apart from MSPs contribute to the gamma-ray emission of these GCs. We also note that variability of NGC 6218 in particular points to a contribution other than MSPs to the gamma-ray emission, possibly a background AGN.  An attempt to correlate the gamma-ray flux with the number of known MSPs in the GCs in our study reinforces this view, though there is some evidence for a correlation between the gamma-ray flux and the mass $\times$ normalised encounter rates.

We note the presence of a link between GC core diffuse X-ray emission and GC gamma-ray emission. The core diffuse X-ray emission could be due to either unresolved point sources or to relativistic electrons in the GCs. In the latter case, one might expect TeV emission from the GCs due to inverse Compton radiation, and observations with ground-based gamma-ray telescopes such as CTA could resolve this issue. 

The link between core diffuse X-ray emission and gamma-ray emission is tentative, largely because there are relatively few gamma-ray emitting GCs for which X-ray observations are available. Further X-ray observations of GCs would be helpful in this regard.

\section*{Acknowledgements}

SJL thanks Jeremy Perkins and Elizabeth Hayes for very helpful advice regarding methods for low energy analysis using \textit{Fermi}-LAT. We acknowledge the excellent data and analysis tools provided by the \textit{Fermi}-LAT collaboration. AMB and PMC acknowledge the financial support of the UK Science and Technology Facilities Council consolidated grant ST/P000541/1. We thank the anonymous referee for comments which helped improve this paper. This research has made use of the SIMBAD database, operated at CDS, Strasbourg, France (\cite{2000A&AS..143....9W}).




\bibliographystyle{mnras}
\bibliography{exportlist} 








\bsp	
\label{lastpage}
\end{document}